\documentclass[12pt]{article} 
\RequirePackage[colorlinks,citecolor=blue,urlcolor=blue]{hyperref}
\usepackage[utf8]{inputenc} 

\usepackage{geometry} 
\usepackage{graphicx} 

\usepackage[dvipsnames]{xcolor}
\usepackage{booktabs} 
\usepackage{array} 
\usepackage{paralist} 
\usepackage{verbatim} 
\usepackage{subfig} 
\usepackage{amsbsy}
\usepackage{amsmath}
\usepackage{amssymb}
\usepackage{amsfonts}
\usepackage{amscd}
\usepackage{amsthm}
\usepackage{float}
\usepackage[toc,page]{appendix}
\usepackage{lineno}
\usepackage{mathrsfs}
\usepackage{lineno}
\usepackage{todonotes}
\usepackage{setspace}
\usepackage{natbib}
\usepackage{multirow}
\usepackage{algorithm,algpseudocode}
\usepackage{varwidth}
\usepackage{ifthen}
\usepackage{pgfplots}
\usepackage{tikz}
\usetikzlibrary{decorations.pathreplacing,arrows,calc,positioning,automata,patterns,arrows.meta,math}

\usepackage{fancyhdr} 
\usepackage{rotating}
\usepackage[group-separator={,},group-minimum-digits=4]{siunitx}

\usepackage{sectsty}
\allsectionsfont{\sffamily\mdseries\upshape} 

\usepackage[nottoc,notlof,notlot]{tocbibind} 
\usepackage[titles,subfigure]{tocloft} 

\graphicspath{{Figures/}}

\tikzset{
    ncbar angle/.initial=90,
    ncbar/.style={
        to path=(\tikztostart)
        -- ($(\tikztostart)!#1!\pgfkeysvalueof{/tikz/ncbar angle}:(\tikztotarget)$)
        -- ($(\tikztotarget)!($(\tikztostart)!#1!\pgfkeysvalueof{/tikz/ncbar angle}:(\tikztotarget)$)!\pgfkeysvalueof{/tikz/ncbar angle}:(\tikztostart)$)
        -- (\tikztotarget)
    },
    ncbar/.default=0.5cm,
}

\tikzset{square left brace/.style={ncbar=0.5cm}}
\tikzset{square right brace/.style={ncbar=-0.5cm}}

\tikzset{round left paren/.style={ncbar=0.5cm,out=120,in=-120}}
\tikzset{round right paren/.style={ncbar=0.5cm,out=60,in=-60}}

\newcommand{\grid}[9]{
  \pgfmathsetmacro{\threadx}{#6}
  \pgfmathsetmacro{\thready}{#7}
  \pgfmathsetmacro{\threadblockx}{#8 + 1}
  \pgfmathsetmacro{\threadblocky}{#9}
  \foreach \x in {1,...,#3} {
    \foreach \y in {1,...,#4} {
      \pgfmathsetmacro{\xone}{#1 + (\x * #5)}
      \pgfmathsetmacro{\yone}{#2 + (\y * #5)}
      \pgfmathsetmacro{\xtwo}{#1 + ((\x + 1) * #5)}
      \pgfmathsetmacro{\ytwo}{#2 + ((\y + 1) * #5)}
      \ifthenelse{\y < \threadblocky \AND \x < \threadblockx } 
      {
        \ifthenelse{\y = \thready \OR \x = \threadx } 
        {
          \draw[draw=gray,line width=0.1mm, fill=Apricot] (\xone, \yone) rectangle (\xtwo, \ytwo);
        }
        {
          \draw[draw=gray,fill=SeaGreen,line width=0.1mm] (\xone, \yone) rectangle (\xtwo, \ytwo);
        }
      }
      {
          \draw[draw=gray,line width=0.1mm, fill=white] (\xone, \yone) rectangle (\xtwo, \ytwo);
      }
    }
  }
	\draw[draw=gray,line width=0.2mm,inner sep=0pt] (#1 + #5,  #2 + #5) rectangle (#1 + #3 * #5 + #5, #2 + #4 * #5 + #5);
}

\newcommand{\gridpartials}[9]{
  \pgfmathsetmacro{\threadx}{#6}
  \pgfmathsetmacro{\thready}{#7}
  \pgfmathsetmacro{\threadblockx}{#8 + 1}
  \pgfmathsetmacro{\threadblocky}{#9 - 1}
  \foreach \x in {1,...,#3} {
  	\pgfmathsetmacro{\xone}{#1 + ((\x) * #5) + (0.3 * (\x- 1)) }
  	\pgfmathsetmacro{\xtwo}{#1 + ((\x + 1) * #5) + (0.3 * (\x - 1))}
		\foreach \y in {1,...,#4} {
       		\pgfmathsetmacro{\yone}{#2  + ((\y) * #5)}
      		\pgfmathsetmacro{\ytwo}{#2 + ((\y + 1) * #5)}
 		    \ifthenelse{\y > \threadblocky \AND \x < \threadblockx } 
	    	 {
	        	\ifthenelse{\y > \thready \AND \x = \threadx } 
	         	{
	          		\draw[draw=gray,line width=0.1mm, fill=Apricot] (\xone, \yone) rectangle (\xtwo, \ytwo);
	        	}
	        	{
	          		\draw[draw=gray,fill=SeaGreen,line width=0.1mm] (\xone, \yone) rectangle (\xtwo, \ytwo);
	        	}
	      	}
	      	{
	        	\draw[draw=gray,fill=white,line width=0.1mm] (\xone, \yone) rectangle (\xtwo, \ytwo);
	      	}
    	}
	  \draw[draw=gray,line width=0.2mm,inner sep=0pt] (\xone, #2 + #5) rectangle (\xtwo, #2 + #4 * #5 + #5);
  }
}

\newcommand{\dotplaceholder}[3]{
  \foreach \x in {0.1,0.2,0.3}{
    \filldraw [black] (#1+\x,#2-\x) circle (0.5pt);
  }
  \node [align=right] at (#1+0.7, #2-0.5) {\scriptsize{#3}};
}

\begin{document}
\begin{flushright}
Article (Methods)\\
Version dated: \today\\
\end{flushright}

\bigskip
\medskip
\begin{center}

\noindent{\Large \bf Many-core algorithms for high-dimensional gradients on phylogenetic trees}
\bigskip

\noindent{\normalsize \sc
	Karthik Gangavarapu$^{1}$, Xiang Ji$^{2}$, Guy Baele$^{3}$, Mathieu Fourment$^{4}$, Philippe Lemey$^{3}$, Frederick A.~Matsen IV$^{5,6,7,8}$ and Marc A.~Suchard$^{1,9,10}$ \\}
\vspace{1em}
\noindent {\small
  \it $^1$Department of Biomathematics, David Geffen School of Medicine at UCLA, University of California,
  Los Angeles, United States \\
  \it $^2$Department of Mathematics, School of Science \& Engineering, Tulane University, New Orleans, United States \\
  \it $^3$Department of Microbiology, Immunology and Transplantation, Rega Institute, KU Leuven, Leuven, Belgium \\
  \it $^4$Australian Institute for Microbiology and Infection, University of Technology Sydney, Ultimo NSW, Australia \\
  \it $^{5}$Public Health Sciences Division, Fred Hutchinson Cancer Research Center, Seattle, Washington, USA\\
  \it $^{6}$Department of Statistics, University of Washington, Seattle, USA\\
  \it $^{7}$Department of Genome Sciences, University of Washington, Seattle, USA\\
  \it $^{8}$Howard Hughes Medical Institute, Fred Hutchinson Cancer Research Center, Seattle, Washington, USA\\
  \it $^{9}$Department of Biostatistics, Jonathan and Karin Fielding School of Public Health, University
  of California, Los Angeles, United States \\
  \it $^{10}$Department of Human Genetics, David Geffen School of Medicine at UCLA, University of California,
  Los Angeles, United States} \\

\end{center}
\medskip
\noindent{\bf Corresponding author:} Marc A.~Suchard, Departments of Biostatistics, Biomathematics, and Human Genetics,
University of California, Los Angeles, 695 Charles E.~Young Dr., South,
Los Angeles, CA 90095-7088, USA; E-mail: \url{msuchard@ucla.edu}

\vspace{1in}

\clearpage

\paragraph{Abstract}

Advancements in high-throughput genomic sequencing are delivering genomic pathogen data at an unprecedented rate, positioning statistical phylogenetics as a critical tool to monitor infectious diseases globally.
This rapid growth spurs the need for efficient inference techniques, such as Hamiltonian Monte Carlo (HMC) in a Bayesian framework, to estimate parameters of these phylogenetic models where the dimensions of the parameters increase with the number of sequences $N$. 
HMC requires repeated calculation of the gradient of the data log-likelihood with respect to (wrt) all branch-length-specific (BLS) parameters that traditionally takes $\mathcal{O}(N^2)$ operations using the standard pruning algorithm.
A recent study proposes an approach to calculate this gradient in $\mathcal{O}(N)$, enabling researchers to take advantage of gradient-based samplers such as HMC.
The CPU implementation of this approach makes the calculation of the gradient computationally tractable for nucleotide-based models but falls short in performance for larger state-space size models, such as Markov-modulated and codon models.
Here, we describe novel massively parallel algorithms to calculate the gradient of the log-likelihood wrt all BLS parameters that take advantage of graphics processing units (GPUs) and result in many fold higher speedups over previous CPU implementations.
We benchmark these GPU algorithms on three computing systems using three evolutionary inference examples exploring complete genomes from 997 dengue viruses, 62 carnivore mitochondria and 49 yeasts, and observe a greater than 128-fold speedup over the CPU implementation for codon-based models and greater than 8-fold speedup for nucleotide-based models.
As a practical demonstration, we also estimate the timing of the first introduction of West Nile virus into the continental Unites States under a codon model with a relaxed molecular clock from 104 full viral genomes, an inference task previously intractable.
We provide an implementation of our GPU algorithms in BEAGLE v4.0.0, an open source library for statistical phylogenetics that enables parallel calculations on multi-core CPUs and GPUs.

\section{Introduction}

\newcommand\nbranches{N}

Genomic sequencing has become a critical tool in monitoring the evolution and spread of infectious pathogens to inform public health interventions, as the unprecedented number of genomes sequenced to monitor the emergence and growth of variants during the ongoing SARS-CoV-2 pandemic demonstrates \citep{OudeMunnink2021,Brito2022}.
This has created the need for statistical phylogenetic methods that can be used to derive useful insights in a timely manner from these large molecular sequence alignments.
Within a Bayesian framework, such analyses typically employ Markov chain Monte Carlo (MCMC) methods such as the random walk Metropolis-Hastings (MH) algorithm \citep{metropolisEquationStateCalculations1953,hastingsMonteCarloSampling} to simultaneously infer the discrete tree topology and continuous branch-length-specific (BLS) parameters such as the branch lengths (or correspondingly, node heights) and branch-specific evolutionary rates.
However, the number of possible tree topologies increases super-exponentially, while the dimensionality of continuous BLS parameters further increases linearly with the number of sequences in the alignment.
There exist state-of-the-art MCMC algorithms such as Hamiltonian Monte Carlo (HMC) that use Hamiltonian dynamics to traverse the parameter space more efficiently compared to a random walk proposal distribution \citep{neal2011mcmc}.
Using HMC enables phylogenetic methods to more efficiently infer continuous high-dimensional BLS parameters, sparing valuable compute time to search the discrete space of possible tree topologies.
Nonetheless, this efficiency comes with the cost of needing to calculate the gradient of the molecular sequence alignment log-likelihood function with respect to (wrt) all BLS parameters, an aspect that remains computationally intensive.

One generally assumes that the individual alignment sites arise conditionally independently from a continuous-time Markov chain (CTMC) process acting along the branches of an estimable evolutionary tree relating the $N$ sequences, where a branch length and, often, an associated evolutionary rate-scalar characterize the branch-specific processes.
Under this CTMC model, Felsenstein's pruning algorithm \citep{felsensteinEvolutionaryTreesDNA1981} renders the calculation of the likelihood  computationally tractable.
The pruning algorithm calculates the probability of only the observed sequence data below each internal node in the tree in a single post-order traversal that visits each node in a descendant-to-parent fashion.
By efficiently reusing these partial likelihood vectors from previously visited nodes, the algorithm reduces the computational complexity of calculating the likelihood to just $\mathcal{O}(N)$.
The same algorithm can be used to calculate the gradient of the log-likelihood wrt a BLS parameter by substituting the CTMC transition probability matrix along one branch with its derivative \citep{kishinoMaximumLikelihoodInference1990,bryant2005likelihood,kenneyHessianCalculationPhylogenetic2012}.
In this manner, calculating the gradient of the log-likelihood wrt all BLS parameters requires $\mathcal{O}(N^2)$ operations.
Until recently, this computational cost has proved too prohibitive for the use of high-dimensional gradient-based samplers or optimizers in statistical phylogenetics.

\cite{ji2020gradients}, however, proposed an algorithm to calculate this gradient in linear-time.
To accomplish this, they introduce a pre-order traversal of the tree that involves visiting each node in a parent-to-descendant fashion after the post-order traversal.
Combining the post- and pre-order partial likelihoods vectors calculated at each node with their branch-specific process derivative yields the whole gradient.
Calculation of the pre-order partial likelihood vectors and their contribution to the gradient is straight-forwardly parallelizable in a multi-core CPU setting by dividing sites in the alignment into conditionally independent partitions that define separate computational tasks.
These calculations may offer further parallelization across different rate categories for models that incorporate among-site rate variation \citep{yangAmongsiteRateVariation1996}.
Despite these existing parallelization schemes on CPUs, the widespread availability of specialized hardware such as graphics processing units (GPUs) opens up the possibility of further speeding up these calculation by many-fold.

GPUs were originally designed for image rendering but have since become ubiquitous for scientific computing.
The availability of application programming interfaces such as OpenCL \citep{stoneOpenCLParallelProgramming2010} and CUDA \citep{cuda} expanded the use of GPUs for general purpose computing beyond graphics processing.
Modern GPUs come equipped with thousands of simple cores, enabling them to carry out a vast number of lock-stepped calculations in parallel.
In contrast, CPUs contain significantly fewer cores, yet each core can perform complex, independent tasks.
This difference allows GPUs to deliver much higher parallelization compared to CPUs for fine-scale numerical operations on large blocks of data.
In the phylogenetic setting, \cite{suchardManycoreAlgorithmsStatistical2009} first described how to calculate the post-order partial likelihood vectors with fine-scale parallelization on GPUs and \cite{ayresBEAGLEImprovedPerformance2019} report recent performance gains.

Building upon this prior work, we present here two novel algorithms that utilize GPUs to calculate the pre-order partial likelihood vectors and the gradient of the log-likelihood wrt all BLS parameters.
We then benchmark computational performance in inferring BLS parameters using our algorithms on the GPU compared to the existing CPU implementation.
We apply our algorithms to date the timing of the first introduction of West Nile virus into the continental United States under a codon model with branch-specific evolutionary rates, an inference task that was previously intractable.
Finally, we highlight the limitations of our current approach and discuss future work to address these limitations and further exploit many-core algorithms for statistical phylogenetics.

\subsection{Considerations for GPU algorithms}

\begin{figure}[H]
  \centering
  \begin{tikzpicture}
    \newcommand\gmx{0}
    \newcommand\gmy{0}
    \filldraw [fill=Dandelion, draw=none] (\gmx,\gmy) rectangle (\gmx+1,\gmy+3.9) node[pos=.5,align=center] {GM};
    \draw[-] (\gmx + 1.5, \gmy+0.6) -- (\gmx + 1.5, \gmy + 2.9);
    \draw[latex-latex,draw=black] (\gmx+1, \gmy+1.95) -- (\gmx+1.5, \gmy+1.95);

    \foreach \y/\smtext in {2.4/SM-1, 0.1/SM-(N)}{
      \filldraw [fill=GreenYellow,draw=none] (\gmx + 1.5 + 0.4, \gmy+\y-0.1) rectangle (\gmx + 5.7,\gmy+\y+1.5);
      \node at (\gmx + 2 + 1.8,\gmy + \y + 1.25) {\scriptsize{\smtext}};


      \filldraw [fill=Cerulean, draw=none] (\gmx + 1.5 + 0.6, \gmy+\y) rectangle (\gmx+1.5 + 0.9 + 1,\gmy+\y+1) node[pos=.5,align=center] {\scriptsize{ShM}};

      \draw[fill=white, draw=none] (\gmx + 4, \gmy+\y) rectangle (\gmx + 5.5, \gmy+\y+1);
      
      \foreach \thready [count=\threadi] in {0.6, 0.3} {
      	      \draw[fill=none](\gmx + 4.1, \gmy+\y+0.1+\thready) rectangle (\gmx + 5.2, \gmy+\y+0.3+\thready) node[pos=0.5] {\tiny{Thread \threadi}};
      	      \draw[fill=gray] (\gmx + 5.2, \gmy+\y+0.1+\thready) rectangle (\gmx + 5.4, \gmy+\y+0.3+\thready);
      }
      
      \draw[densely dotted,line width=0.75,black] (\gmx + 4.75, \gmy + \y + 0.3) -- (\gmx + 4.75, \gmy + \y+0.075);

      \draw[-latex,draw=black] (\gmx+1.5, \gmy+\y +0.5) -- (\gmx+2.1, \gmy+\y+0.5);
      \draw[latex-latex, draw=black] (\gmx+3.4, \gmy+\y+0.5) -- (\gmx+4, \gmy+\y+0.5);
    }

    \draw[densely dotted,line width=0.75,black] (\gmx + 3.7, \gmy + 2.15) -- (\gmx + 3.7, \gmy + 1.75);

  \end{tikzpicture}
  \caption{GPU memory hierarchy. Each GPU is equipped with hundreds to thousands of parallel processors known as streaming multiprocessors (SMs) shown in green. Each SM can execute multiple threads in parallel and threads are further grouped into thread-blocks. Each thread has exclusive access to register memory shown in gray that other threads cannot access. All threads within a thread-block can access the same on-chip shared memory (ShM) shown in blue. GPUs are also equipped with global memory (GM) shown in orange which can be accessed by thread-blocks across all SMs.}
  \label{fig:gpumemory}
\end{figure}
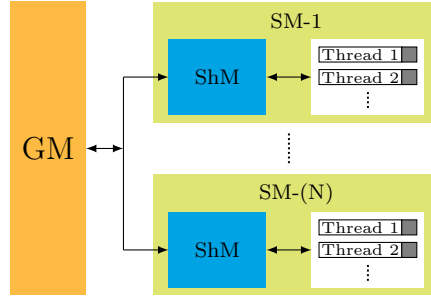

The design of our algorithms is guided by certain aspects of GPU architecture to achieve optimal performance on the available resources.
A GPU can consist of hundreds to over a thousand or so parallel processors known as streaming multiprocessors (SMs) or CUDA cores on NVIDIA devices.
Each SM can execute multiple threads that are further grouped into thread-blocks with each thread-block containing up to 512 to 1024 threads as determined by hardware constraints.
Within a thread-bock, groups of 16 or 32 consecutive threads, termed warps, are executed in parallel using a single instruction, multiple threads (SIMT) execution model.
Functions, called kernels, are executed by thread-blocks that are indexed as a grid.
Each SM has 256 KB of register memory (64,000 32-bit registers) and each thread can be allocated a maximum of 1 KB (255 registers).
Registers are allocated exclusively to a single thread and cannot be accessed by other threads. 
In addition to register memory, all threads within a thread-block are allocated shared memory (ShM) of 32 or 64 KB as determined by hardware constraints and all thread-blocks have access to the global memory (GM) on the GPU that is typically a dynamic random-access memory device (Figure \ref{fig:gpumemory}).
ShM is located on chip and, hence, memory access can be 100- to 150-fold faster than GM transactions.
Since all threads within a thread-block have access to the same ShM, each thread within a block can cooperatively fetch values from global memory and cache them in shared memory for subsequent operations.
To further minimize the number of memory transactions with GM, memory access by consecutive threads within a warp or half warp, are combined or `coalesced' into one or more 32-, 64- or 128-byte transactions based on hardware constraints.
The most optimal GM access pattern is achieved when consecutive threads within a warp or half warp access consecutive memory addresses in GM which maximizes the use of the available bandwidth of each `coalesced memory transaction'.

Once a warp executes an instruction, it has to wait for a fixed number of clock cycles, called the latency, before executing the next instruction.
Typically the latency for an instruction that requires GM access is hundreds of clock cycles compared to far fewer clock cycles for ShM access.
Maximum utilization of available resources on the GPU can be achieved by ensuring that there are sufficient warps queued to execute a new instruction at each clock cycle while other warps wait until their memory transactions are complete.
This approach hides memory latency and motivates the fine-scale parallelization of any numerical operation executed using a GPU. 
In addition to having low latency ShM, synchronization between threads within the same block i.e, coordinating the execution of threads when they access the same resources such as ShM or GM is inexpensive. 
This makes parallelization efficient for even small binary reductions of numbers which is not the case for multi-core CPUs.

\section{Methods}

\newcommand\data{\mathbf{Y}}
\newcommand{\datum}{Y}
\newcommand\columns{C}
\newcommand\column{c}
\newcommand\states{S}
\newcommand\state{s}
\newcommand\numtips{N}
\newcommand\branchLengths{T}
\newcommand\branchLength{b}
\newcommand\terminalBranches{\varepsilon}
\newcommand\internalBranches{\mathcal{I}}
\newcommand\phylogeny{{\cal F}}
\newcommand\branch{b}
\newcommand\rates{R}
\newcommand\rate{r}
\newcommand{\siteSpecificRate}{\gamma}
\newcommand\postorderPartial{\mathbf{p}}
\newcommand\postorderElement{p}
\newcommand\preorderPartial{\mathbf{q}}
\newcommand\preorderElement{q}
\newcommand{\probabilityMatrix}[2]{\mathbf{P}^{(#1)} \hspace{-0.2em} \left( #2 \right)}
\newcommand{\probabilityEntry}[4]{P^{(#1)}_{#3 #4} \hspace{-0.2em} \left( #2 \right)}
\newcommand\rateMatrix{\mathbf{Q}}
\newcommand\disjointSet{Y}
\newcommand{\below}[1]{\mathbf{Y}_{\lfloor #1 \rfloor}}
\newcommand{\abbove}[1]{\mathbf{Y}_{\lceil #1 \rceil}}
\newcommand{\transpose}{'}
\newcommand{\estate}{t}
\newcommand{\parent}{k}
\newcommand{\sibling}{j}
\newcommand\currnode{i}
\newcommand{\probability}[1]{\mathbb P \hspace{-0.2em} \left( #1 \right)}
\newcommand{\rootDistribution}{\boldsymbol{\pi}}

The molecular sequence alignment $\data= \left( \data_1,\ldots,\data_{\columns} \right)$ comprises $\columns$ aligned columns or sites, where column data $\data_c = \left( \datum_{1c}, \ldots, \datum_{Nc} \right)\transpose$ for $\column = 1,\ldots,\columns$ contains one homologous sequence character for each of the $\numtips$ taxa.
Following standard practice since \cite{felsensteinEvolutionaryTreesDNA1981}, we assume that $\data_c$ are conditionally independent and identically distributed.
Thus, it suffices to compute the post- and pre-order partial likelihood vectors and gradient using only the unique $\data_c$ and reweigh these quantities appropriately using standard data compression techniques.
Each aligned character $\datum_{nc}$ for $n = 1, \ldots, \numtips$ exists in one of $\states$ possible states that we arbitrarily label $\left\{ 1,\ldots,\states \right\}$.
For nucleotide alignments, the state-space size $\states = 4$; likewise amino acid and codon alignments yield $\states = 20$ and $\states = 61$, respectively.
Additionally, for phylogeographic inference, $\states$ can be large, often comparable to codon models \citep{dudasVirusGenomesReveal2017a,lemeyAccommodatingIndividualTravel2020a} and has eclipsed 200 in Markov-modulated CTMC models \citep{baeleMarkovModulatedContinuousTimeMarkov2021}.
We observe these characters from $\numtips$ taxa related by an (often) unknown phylogeny $\phylogeny$.
This phylogeny is a directed, bifurcating graph with $\numtips$ tip nodes corresponding to the taxa that we label $\left( 1,\ldots,\numtips \right)$, $\numtips - 2$ internal nodes that we label $\left( \numtips+1,\ldots,2\numtips-2 \right)$ and one root node that we label $2\numtips-1$.
Connecting each parent node to its child in $\phylogeny$ are $2\numtips - 2$ edges or branches with branch lengths $\left( \branchLength_1,\ldots,\branchLength_{2\numtips-2} \right)$ that we index via their child node number.
Each branch length $\branchLength_i$ for $i = 1, \ldots, 2 \numtips - 2$ can measure the expected number of character substitutions along that branch or be the difference between the parent and child node heights measured in time-units multiplied by a (possibly branch-specific) evolutionary rate scalar.
Without loss of generality, unrooted phylogenies are nested within this formulation by setting one of the branch lengths emerging from the root to zero.

An $\states \times \states$ infinitesimal generator matrix $\rateMatrix$ characterizes the CTMC process.
Into this formulation, we further incorporate site-specific rate variation using the popular discretized models \citep{yangMaximumLikelihoodPhylogenetic1994} that modulate the CTMC for each column independently through
a finite mixture of rate categories $\rate = \left\{1, \ldots, R\right\}$ where each category corresponds to an overall rate scale $\siteSpecificRate_r$ drawn with probability
$\probability{\siteSpecificRate_r}$.
Thus under rate category $\rate$, along each branch $\currnode$ in $\phylogeny$, the CTMC posits that substitutions arise according to an $\states \times \states$ finite-time transition probability matrix $\probabilityMatrix{r}{\branchLength_{\currnode}} = \left\{
\probabilityEntry{r}{\branchLength_{\currnode}}{\state}{\estate} \right\}$, where
\begin{equation}
\probabilityMatrix{r}{\branchLength_{\currnode}} = \exp \left( \siteSpecificRate_{\rate} \branchLength_{\currnode} \rateMatrix \right) ,
\end{equation}
such that the $\state \estate^{\text{\tiny th}}$ element $\probabilityEntry{r}{\branchLength_{\currnode}}{\state}{\estate}$ is the probability of observed or unobserved state $\estate$ at the child node of branch $\currnode$ given observed or unobserved state $\state$ at the parent node.

To form the partial likelihood vectors calculated during the post- and then pre-order traversals and understand how they relate to the sequence likelihood and then its gradient, we require some data augmentation.
For column $\column$, let $\datum_{ic}$ for $i = \numtips + 1, \ldots, 2 \numtips - 1$ represent the unobserved (latent) character states at the internal and root nodes.
Further, we can divide the observed characters $\data_{\column}$ at the tips into two disjoint sets wrt to any node in $\phylogeny$.
Let $\below{\currnode \column}$ be the observed characters at the tips descendant of node $\currnode$, noting that $\below{2 \numtips - 1,\column} = \data_{\column}$, and let $\abbove{\currnode \column} = \data_{\column} / \below{\currnode \column}$ denote the observed characters at the tips not descendant from node $\currnode$.

The post-order partial likelihood vector is denoted by $\postorderPartial_{\currnode \rate \column} = (\postorderElement_{\currnode \rate \column 1}, \ldots, \postorderElement_{\currnode \rate \column S})\transpose$, where
$\postorderElement_{\currnode \rate \column \state} = \probability{\below{\currnode \column} | \datum_{\currnode \column} = \state, \siteSpecificRate_r}$ at node $\currnode$ under rate category $\rate$ for column $\column$.
We compute these vectors using Felsenstein's pruning algorithm via recursive application of
\begin{equation}
	\postorderPartial_{\parent \rate \column} =
	\left[ \probabilityMatrix{r}{\branchLength_{\currnode}}\right] \postorderPartial_{\currnode \rate \column}
	\circ
	\left[\probabilityMatrix{r}{\branchLength_{\sibling}}  \right] \postorderPartial_{\sibling \rate \column} ,
\end{equation}
where node $\parent$ is parent to node $\currnode$ and its sibling node $\sibling$ and $\circ$ signifies component-wise multiplication.
If we let $\rootDistribution = \left( \probability{Y_{2\numtips -1,c} = 1}, \ldots ,\probability{Y_{2\numtips-1,c}=\states} \right)\transpose$ denote an arbitrary prior state distribution vector at the root that is often set equal to the stationary distribution of $\rateMatrix$, then
\begin{align}
	\probability{\data} &= \prod_{\column = 1}^{\columns} \sum_{\rate = 1}^{\rates} \probability{\data_{\column} | \siteSpecificRate} \probability{\siteSpecificRate_r} \nonumber \\
	&= \prod_{\column = 1}^{\columns} \sum_{\rate = 1}^{\rates}
		\left[
			\postorderPartial_{2 \numtips - 1, \rate \column}\transpose \, \rootDistribution
		\right]
		\probability{\siteSpecificRate_r}
\end{align}
yields the sequence likelihood.
\cite{suchardManycoreAlgorithmsStatistical2009} develop massively parallel algorithms for calculating matrices $\probabilityMatrix{\rate}{\branchLength_{\currnode}}$ for all $\currnode$ and $\rate$ simultaneously and vectors $\postorderPartial_{\currnode \rate \column}$ for all $\rate$ and $\column$ simultaneously on GPUs.
Figure \ref{fig:postOrderKernel} depicts the parallel thread-block design for the algorithm to calculate the post-order partial likelihood vectors described in \cite{suchardManycoreAlgorithmsStatistical2009}.

\cite{ji2020gradients} introduce the pre-order partial likelihood vector  $\preorderPartial_{\currnode \rate \column} =
\left(
	\preorderElement_{\currnode \rate \column 1},
	\ldots ,
	\preorderElement_{\currnode \rate \column \states}
\right)\transpose$,
where
$\preorderElement_{\currnode \rate \column \state} =
\probability{\datum_{\currnode \column} = \state, \abbove{\currnode \column}}$ at node $\currnode$ for column $\column$ under rate category $\rate$ and demonstrate how to compute these vectors recursively given the post-order partial likelihood vectors and transition matrices.
Starting from the root and assigning $\preorderPartial_{2 \numtips - 1, \rate \column} = \rootDistribution$, we continue toward the tips via
\begin{equation}
	\preorderPartial_{\currnode \rate \column} =
	\left[ \probabilityMatrix{r}{\branchLength_{\currnode}} \right]\transpose
	\left\{
		\preorderPartial_{\parent \rate \column}
		\circ
		\left[ \probabilityMatrix{r}{\branchLength_{\sibling}} \right]
		\postorderPartial_{\sibling \rate \column}
	\right\} .
	\label{eq:preorder}
\end{equation}

The value of these pre-order partial likelihood vectors shines in the realization that
\begin{align}
\probability{\data}
	&= \prod_{\column = 1}^{\columns} \sum_{\rate = 1}^{\rates}
		\left[
			\postorderPartial_{\currnode \rate \column}\transpose
			\preorderPartial_{\currnode \rate \column}
		\right]
		\probability{\siteSpecificRate_r} \text{ for \textit{any} node $\currnode$} ,
\end{align}
and this insight enables a linear-in-$\numtips$-time approach to evaluate the gradient of the log-likelihood wrt all BLS parameters
\begin{align}
\nabla \log \probability{\data} =
\left(
	\frac{\partial}{\partial \branchLength_{1}} \log \probability{\data},
	\ldots,
	\frac{\partial}{\partial \branchLength_{2\numtips - 2}} \log \probability{\data}
\right)\transpose ,
\end{align}
because
\begin{align}
\frac{\partial}{\partial \branchLength_{\currnode}} \log \probability{\data} &=
\sum_{\column = 1}^{\columns}
		\frac{\partial}{\partial \branchLength_{\currnode}} \log \probability{\data_{\column}}
	 \label{eq:gradientSumAcrossColumns}
\end{align}
and
\begin{align}
\frac{\partial}{\partial \branchLength_{\currnode}} \log \probability{\data_{\column}}
	&=
	\frac{\partial}{\partial \branchLength_{\currnode}} \log
	\left(
		\sum_{\rate = 1}^{\rates}
		\left[
			\postorderPartial_{\currnode \rate \column}\transpose
			\preorderPartial_{\currnode \rate \column}
		\right]
		\probability{\siteSpecificRate_r}
	\right) \nonumber \\
	&=
	\frac{
		\sum_{\rate = 1}^{\rates}
		\frac{\partial}{\partial \branchLength_{\currnode}}
		\left[
			\postorderPartial_{\currnode \rate \column}\transpose
			\preorderPartial_{\currnode \rate \column}
		\right]
		\probability{\siteSpecificRate_r}
	}{
		\sum_{\rate = 1}^{\rates}
		\left[
			\postorderPartial_{\currnode \rate \column}\transpose
			\preorderPartial_{\currnode \rate \column}
		\right]
		\probability{\siteSpecificRate_r}
	}
	\nonumber \\
	&=
	\frac{
		\sum_{\rate = 1}^{\rates}
		\left[
			\postorderPartial_{\currnode \rate \column}\transpose
			\frac{\partial}{\partial \branchLength_{\currnode}}
			\preorderPartial_{\currnode \rate \column}
		\right]
		\probability{\siteSpecificRate_r}
	}{
		\sum_{\rate = 1}^{\rates}
		\left[
			\postorderPartial_{\currnode \rate \column}\transpose
			\preorderPartial_{\currnode \rate \column}
		\right]
		\probability{\siteSpecificRate_r}
	}
	\nonumber \\
	&=
	\frac{
		\sum_{\rate = 1}^{\rates}
		\siteSpecificRate_{\rate}
		\left[
			\postorderPartial_{\currnode \rate \column}\transpose
			\rateMatrix\transpose
			\preorderPartial_{\currnode \rate \column}
		\right]
		\probability{\siteSpecificRate_r}
	}{
		\sum_{\rate = 1}^{\rates}
		\left[
			\postorderPartial_{\currnode \rate \column}\transpose
			\preorderPartial_{\currnode \rate \column}
		\right]
		\probability{\siteSpecificRate_r}
	}
	 ,
	\label{eq:gradient}
\end{align}
or, more intuitively, all elements in the gradient are simple weighted sums of weighted inner products involving $\postorderPartial_{\currnode \rate \column}$ and $\preorderPartial_{\currnode \rate \column}$.

\subsection{Computing the pre-order partial likelihood vectors}

Algorithm \ref{alg:preorderPartial} outlines our approach to massively parallelize the computation of Equation \ref{eq:preorder} across all $r$ and $c$ simultaneously on a GPU, with each $(\rate,\column,\state)$-entry processed in its own short-lived thread.
Naturally, this algorithm builds on the success of the post-order partial likelihood algorithm of \citet{suchardManycoreAlgorithmsStatistical2009} with several important differences.
First, lines 7 - 10 and 13 - 16 split the matrix-vector multiplications into two serial operations to satisfy the dependencies of Equation \ref{eq:preorder}.
Threads within a thread-block start executing operations on $\probabilityMatrix{\rate}{\branchLength_{\currnode}}\transpose$ only after completing operations involving $\probabilityMatrix{\rate}{\branchLength_{\sibling}}$.
While this reduces instruction-level parallelism, it removes the burden of storing both matrices in ShM simultaneously.
The second important difference is that Equation \ref{eq:preorder} requires the transpose of $\probabilityMatrix{\rate}{\branchLength_{\currnode}}$.
When the state-space size is small as is the case for nucleotide models with $\text{\states} = 4$, we can calculate the transpose of the matrix while reading it from GM and comfortably hold all 16 elements of the matrix in ShM.
However, for models with a large state-space size such as codon models with $\states=61$, all $\states^2$ transition probabilities do not fit in ShM simultaneously.
For such cases, we also develop a `matrixTranspose' kernel to calculate the transpose of $\probabilityMatrix{\rate}{\branchLength_{\currnode}}$ in a parallelized manner using the GPU according to Algorithm \ref{alg:matrixtranspose} (Figure \ref{fig:matrixtranspose}).
The resulting transposed matrix is stored in GM and subsequently used to calculate the pre-order partial likelihood vector $\preorderPartial_{\currnode \rate \column}$.

\begin{algorithm}
  \caption{GPU-based parallel computation of pre-order partial likelihood vectors}\label{alg:preorderPartial}
  \begin{algorithmic}[1]
    \State \textbf{define} COLUMN\_BLOCK\_SIZE (CBS) $=$ number of patterns processed in parallel per thread-block
    \State \textbf{define} PEELING\_BLOCK\_SIZE (PBS) $=$ number of states processed in parallel per inner-loop
    \For {\textbf{all} thread-blocks (rate category $r = 1,...,R$ and column-block $= 1,..., \lceil C / \text{CBS} \rceil$) \textbf{in parallel}}
      \For
      	{\textbf{all} threads in block (state $s=1,...,S$ and column $c=1,..., \text{CBS}$) \textbf{in parallel}}
      \State
      	\begin{varwidth}[t]{\linewidth-3em} 
	     \textbf{prefetch} post-order partial likelihood elements $p_{{\sibling}rcs}$ and pre-order partial likelihood elements $q_{{\parent}rcs}$ for CBS columns (reused by all threads in block) where node $\parent$ is parent to node $\currnode$ and its sibling node $\sibling$.
		\end{varwidth}
      \State \textbf{initialize} $\phi$ $\gets$ 0
      \For {$\estate = 1,\ldots,\states$ in PBS-sized \textbf{parallel} chunks}
        \State
        	\begin{varwidth}[t]{\linewidth-5em}
		        \textbf{prefetch} transition probability elements
		        $\probabilityEntry{r}{\branchLength_{\sibling}}{\state}{\estate}$
		        for PBS states (reused by all threads in the block) and \textbf{synchronize}
		    \end{varwidth}
        \State \textbf{increment} $\phi \gets \phi + \probabilityEntry{r}{\branchLength_{\sibling}}{\state}{\estate} \times p_{{\sibling}rc{\estate}}$
      \EndFor
      \State \textbf{form} component-wise product $ q_{{\parent}rcs}
      \gets q_{{\parent}rcs} \times \phi $
      \State \textbf{initialize} $\omega$ $\gets$ 0
      \For {$\estate = 1,\ldots,\states$ in PBS-sized \textbf{parallel} chunks}
        \State
        	\begin{varwidth}[t]{\linewidth-5em}
        		\textbf{prefetch} (transposed) transition probability
        		$\probabilityEntry{r}{\branchLength_{\currnode}}{\estate}{\state}$
        		for PBS states (reused by all threads in the block) and \textbf{synchronize}
        	\end{varwidth}
        \State \textbf{increment} $\omega \gets \omega +
        q_{{\parent}rct} \times \probabilityEntry{r}{\branchLength_{\currnode}}{\estate}{\state}$
      \EndFor
      \State \Return $\omega$ as $ q_{{\currnode}rcs}$
    \EndFor
    \EndFor
  \end{algorithmic}
\end{algorithm}

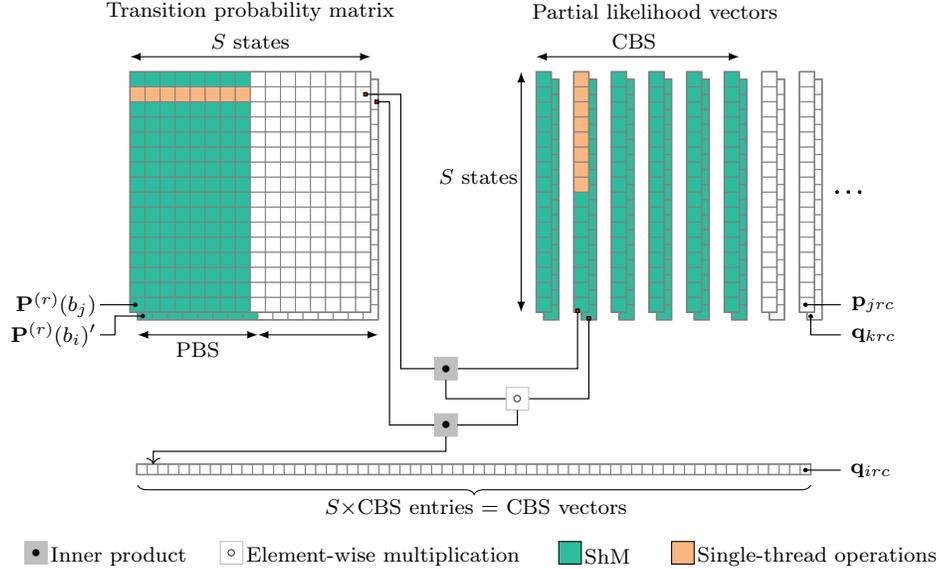
\begin{figure}[H]
  \centering
  \newcommand\Qx{0}
  \newcommand\Qy{3}
  \newcommand\cellwidth{0.2}
  \newcommand\cellwidthtwo{0.14}
  \begin{tikzpicture}
    \node at (\Qx + 0.3 + 1.6,\Qy+4.3) {\scriptsize{Transition probability matrix}};
    \node at (\Qx + 0.3 + 1.6,\Qy+3.9) {\scriptsize{$S~\text{states}$}};
    \draw [latex-latex] (\Qx + 0.3,\Qy+3.7) -- (\Qx + 0.3 + 3.2,\Qy+3.7);

    \grid{\Qx + 0.2}{\Qy}{16}{16}{\cellwidth}{-1}{15}{8}{17}
    \grid{\Qx + 0.1}{\Qy+ 0.1}{16}{16}{\cellwidth}{-1}{15}{8}{17}

    \node at (\Qx + 0.4 + 0.8,  \Qy - 0.2) {\scriptsize{PBS}};
  	\draw [latex-latex] (\Qx + 0.4, \Qy) -- (\Qx+0.4 + 1.6, \Qy);
  	\draw [latex-latex] (\Qx + 0.4 + 1.6, \Qy) -- (\Qx+0.4 + 3.2, \Qy );

    \tikzmath{
		\tmlegendxone  = \Qx;
		\tmlegendyone = \Qy+0.4;
		\tmlegendxtwo = \Qx;
		\tmlegendytwo = \Qy;
	}

    \node [left] at (\tmlegendxone, \tmlegendyone) {\scriptsize{
    	$\probabilityMatrix{r}{\branchLength_{\sibling}}$
    }};

    \node [left] at (\tmlegendxtwo, \tmlegendytwo) {\scriptsize{
	    $\probabilityMatrix{r}{\branchLength_{\currnode}}\transpose$
    }};

    \draw[{Circle[width=0.6mm,length=0.6mm]}-,line width=0.15mm] (\tmlegendxone+0.4, \tmlegendyone) -- (\tmlegendxone-0.1, \tmlegendyone);
    \draw[{Circle[width=0.6mm,length=0.6mm]}-,line width=0.15mm] (\tmlegendxtwo+0.5, \tmlegendytwo+0.25) -- (\tmlegendxtwo+0.1, \tmlegendytwo+0.25) -- (\tmlegendxtwo+0.1, \tmlegendytwo) -- (\tmlegendxtwo-0.1, \tmlegendytwo);

    \newcommand\px{5.5}
    \newcommand\py{3.1}

    \gridpartials{\px+0.1}{\py-0.1}{8}{16}{\cellwidth}{-1}{-1}{6}{-1}
    \gridpartials{\px}{\py}{8}{16}{\cellwidth}{2}{8}{6}{-1}

	\tikzmath{
		\pvlegendxone  =\px+4.15;
		\pvlegendyone = \tmlegendyone;
		\pvlegendxtwo = \px+4.15;
		\pvlegendytwo = \tmlegendytwo;
	}

    \node [right] at (\pvlegendxone, \pvlegendyone) {\scriptsize{
    	$\postorderPartial_{\sibling \rate \column}$

    }};
    \node [right] at (\pvlegendxtwo, \pvlegendytwo) {\scriptsize{
    	$\preorderPartial_{\parent \rate \column}$
    }};

    \draw[{Circle[width=0.6mm,length=0.6mm]}-,line width=0.15mm] (\pvlegendxone - 0.4, \pvlegendyone) -- (\pvlegendxone+0.1, \pvlegendyone);
    \draw[{Circle[width=0.6mm,length=0.6mm]}-,line width=0.15mm] (\pvlegendxtwo-0.3, \pvlegendytwo + 0.275) -- (\pvlegendxtwo-0.3, \pvlegendytwo) -- (\pvlegendxtwo+0.1, \pvlegendytwo);

    \node [right] at (\px,\py+4.2) {\scriptsize{Partial likelihood vectors}};

    \draw [latex-latex] (\px + 0.2, \py+3.6) -- (\px+2.9, \py + 3.6) node [black,midway, below] {};
    \node [] at (\px+1.5, \py+3.8) {\scriptsize{CBS}};

	\node[left] at (\px+0.2,\py+2) {\scriptsize{$\states~\text{states }$}};
	\draw [latex-latex] (\px,\py+0.2) -- (\px,\py+3.4);

    \foreach \x in {0.1,0.25,0.4}{
    	\filldraw [black] (\px+4.1+\x,\py+1.8) circle (0.5pt);
    }

    \newcommand\ox{0.25}
    \newcommand\oy{1}


    \grid{\ox}{\oy}{64}{1}{\cellwidthtwo}{2}{-1}{-1}{2}

    \draw [decorate,decoration={brace,amplitude=3.5pt}]
      (\ox+9.1, \oy) -- (\ox+0.15, \oy) node [black,midway] {};
    \node [] at (\ox+4.65, \oy-0.3) {\scriptsize{$S \times $\text{CBS} entries $=$ CBS vectors}};

    \newcommand\wx{\Qx+1.6}
    \newcommand\wy{\Qy-1}

    \tikzmath{
    	\oponex = \Qx+4.5;
    	\oponey = \wy+0.55;
    	\opwidth = 0.15;
	   	\tmxone = \Qx + 0.55 + 3.35;
		\tmyone = \Qy + 0.35 + 2.85;
		\tmxtwo = \Qx + 4.2 + 0.15;
		\tmytwo = \wy + 0.55;
		\pvxone = \px + 0.75 ;
		\pvyone = \py + 0.25;
		\pvxtwo = \Qx + 4.65;
		\pvytwo = \wy+0.55;
 		\outxone= \Qx+4.35+0.15;
 		\outyone = \wy + 0.4;
 		\outxtwo = \ox+0.21 ;
 		\outytwo = \oy+0.1;
    }

    \draw[{Square[width=0.6mm,length=0.6mm,fill=red]}-,line width=0.15mm] (\pvxone, \pvyone) --  (\pvxone,\pvytwo) -- (\pvxtwo, \pvytwo);
    \draw [draw=lightgray, fill=lightgray] (\oponex-\opwidth,\oponey-\opwidth) rectangle (\oponex+\opwidth,\oponey+\opwidth) node[pos=0.5] {\tiny{$\bullet$}};
  	\draw[{Square[width=0.6mm,length=0.6mm,fill=red]}-,line width=0.15mm] (\tmxone-0.5, \tmyone) -- (\tmxone, \tmyone)  -- (\tmxone, \tmytwo) -- (\tmxtwo, \tmytwo);

    \draw[{Square[width=0.6mm,length=0.6mm,fill=red]}-,line width=0.15mm] (\pvxone + 0.15, \pvyone - 0.1) --  (\pvxone+0.15,\pvytwo-0.4) -- (\pvxtwo + 0.95, \pvytwo-0.4);
	\draw[line width=0.15mm] (\outxone, \outyone) -- (\outxone, \pvytwo-0.4) -- (\pvxtwo + 0.65, \pvytwo-0.4);
	\draw [draw=lightgray] (\pvxtwo + 0.8 - \opwidth, \pvytwo - 0.4 - \opwidth) rectangle (\pvxtwo + 0.8 + \opwidth, \pvytwo - 0.4+\opwidth) node[pos=0.5] {\tiny{$\circ$}};

	\draw[{Square[width=0.6mm,length=0.6mm,fill=red]}-,line width=0.15mm] (\tmxone - 0.35, \tmyone - 0.1) -- (\tmxone - 0.15, \tmyone - 0.1)  -- (\tmxone - 0.15, \tmytwo -0.75) -- (\tmxtwo, \tmytwo-0.75);
	\draw [draw=lightgray, fill=lightgray] (\oponex-\opwidth,\oponey-0.75-\opwidth) rectangle (\oponex+\opwidth,\oponey-0.75+\opwidth) node[pos=0.5] {\tiny{$\bullet$}};
	\draw[line width=0.15mm] (\oponex+\opwidth,\oponey-0.75) -- (\pvxtwo + 0.8,\oponey-0.75) -- (\pvxtwo + 0.8, \pvytwo - 0.4 -\opwidth);

	\draw[->,line width=0.15mm] (\oponex,\oponey-0.75-\opwidth) -- (\oponex,\oponey-1.1) -- (\outxtwo+0.15, \oponey-1.1) -- (\outxtwo+0.15, \oponey-1.275);

    \draw[{Circle[width=0.6mm,length=0.6mm]}-,line width=0.15mm] (\ox+9,\oy+0.2) -- (\ox+9.5,\oy+0.2);
    \draw [draw=none] (\ox+9.4,\oy-0.1) rectangle (\ox+10.1,\oy+0.5) node[pos=0.5] {
    	\scriptsize{\, \, $\preorderPartial_{\currnode \rate \column}$}
    };

    \draw [draw=lightgray, fill=lightgray] (\Qx-1.1, \Qy-2.75) rectangle (\Qx-0.8, \Qy-3.05) node[pos=0.5] {\tiny{$\bullet$}};
    \node [right] at (\Qx-0.9, \Qy-2.95) {\scriptsize{Inner product}};

    \draw [draw=lightgray] (\Qx+1.5, \Qy-2.75) rectangle (\Qx+1.8, \Qy-3.05) node[pos=0.5] {\tiny{$\circ$}};
    \node [right] at (\Qx + 1.7, \Qy-2.95) {\scriptsize{Element-wise multiplication}};

	\draw [fill=SeaGreen] (\Qx+6, \Qy-2.75) rectangle (\Qx+6.3, \Qy-3.05);
	\node [right] at (\Qx + 6.2, \Qy-2.95) {\scriptsize{ShM}};

	\draw [fill=Apricot] (\Qx+7.5, \Qy-2.75) rectangle (\Qx+7.8, \Qy-3.05);
	\node [right] at (\Qx + 7.7, \Qy-2.95) {\scriptsize{Single-thread operations}};

  \end{tikzpicture}
  \caption{Parallel thread-block design to compute pre-order partial likelihood vectors $\preorderPartial_{\currnode \rate \column}$.
  One block evaluates column block size $(\text{CBS}) \times \states$ entries in parallel and prefetches pruning block size $(\text{PBS}) \times \states$ transition probability entries at a time within an inner serial loop (Algorithm \ref{alg:preorderPartial}: Lines 7 - 10 and  13 - 16). Entries fetched from global memory (GM) into shared memory (ShM) are indicated in green and an instance of a single-thread operation is shown in orange.}
   \label{fig:preOrderKernel}
\end{figure}

Per thread, a surprisingly small portion of the code is dedicated to actually compute $\preorderPartial_{irc}$.
Most of the work involves efficiently fetching the transition probabilities and pre- and post-order partial likelihood vectors from GM into ShM to be reused by threads within a block.
Figure \ref{fig:preOrderKernel} outlines the parallel thread-block design for Algorithm \ref{alg:preorderPartial}.
We observe in Equation \ref{eq:preorder} that all $\states$ partials in $q_{\currnode \rate \column}$ for a column $\column$ under a rate class $\rate$, depend on the same partial likelihood vectors $\preorderPartial_{\parent \rate \column}$ and $\postorderPartial_{\sibling \rate \column}$.
Consequentially, we construct $\rates \times \lceil C/\text{CBS} \rceil$ thread blocks, where $\lceil . \rceil$ is the ceiling function and column-block size (CBS) is a design constant that controls the number of columns processed in a block.
Each thread-block shares $\states \times \text{CBS}$ threads that correspond to all $\states$ states for $\text{CBS}$ columns.
All $\states \times \text{CBS}$ threads within a block cooperatively fetch $\states$-lengthed vectors $\preorderPartial_{\parent \rate \column}$ and $\postorderPartial_{\parent \rate \column}$ for CBS columns.
Equation \ref{eq:preorder} also shows that under a rate class $r$, $\preorderPartial_{\currnode \rate \column}$ for all columns $\column \in \columns$ depends on the same finite-time transition probability matrices, $\probabilityMatrix{\rate}{\branchLength_{\currnode}}$ and $\probabilityMatrix{\rate}{\branchLength_{\sibling}}$.
Hence, we use $\states$ threads to fetch columns from each of these matrices which are then reused by all threads in the thread-block.
Each thread-block computes CBS pre-order partial likelihood vectors each with $\states$ partials.

To coalesce GM transactions and efficiently utilize the available memory bandwidth, we ensure that consecutive threads within a block attempt to access consecutive memory addresses only in multiples of 16 values at a time.
For models with a state size $\states$ that is not a multiple of 16, we embed the transition matrices and the partial likelihood vectors into a larger space by adding zeroes as extra entries.
This approach is called `padding' and ensures optimal utilization of the memory bandwidth. 
For example, for codon models with $\states = 61$, we pad the $\states$ with three zero entries yielding a state-space size of 64.
For nucleotide models with $\states = 4$, each thread simply processes four columns of the alignment instead of one.

Since each thread-block can have up to 512 threads (or 1024 threads depending on the hardware), we set CBS to be as large as possible such that $\states \times \text{CBS} \le 512$ without overflowing ShM on the device.
For nucleotide models, we set $\text{CBS}=16$ with each thread processing 4 columns.
An additional complication arises for models with large state-space sizes such as codon models wherein all $\states^2$ transition probabilities do not fit in ShM.
In such cases, $\states$ threads within a thread-block cooperatively fetch columns of the matrix in peeling-block size (PBS) length chunks from GM into ShM.
Thus, for codon models, we set $\text{CBS}=8$ along with an additional design constant, $\text{PBS}=8$.
To ensure that the GM reads of the matrix columns are coalesced, we exploit a column-wise flattened representation of the finite-time transition probability matrices which differs from the standard row-wise representation in modern computing.
While newer GPUs have larger ShM available and can accommodate larger values of PBS and CBS, we set these design constraints to ensure compatibility across a broad range of devices.

\newcommand{\ttab}{\hspace{2em}}

\subsection{Computing the gradient of the log-likelihood wrt all branch-specific parameters}

Algorithm \ref{alg:gradient} outlines our implementation of Equation \ref{eq:gradient}  to calculate the column-specific contribution to the gradient of the log-likelihood wrt all BLS parameters.
To do this, we reuse the pre-order partial likelihood vectors calculated using Algorithm \ref{alg:preorderPartial} and the post-order partial likelihood vectors calculated using the algorithm described in \cite{suchardManycoreAlgorithmsStatistical2009}.
When a post-order partial likelihood vector $\postorderPartial_{\currnode \rate \column}$ points to a tip node and $\datum_{ic}$ is directly observed, researchers often store this vector in a compressed format that entails a single integer (or smaller) filled with the character state to reduce memory access demands.
So, for the tip nodes we simply read in the integer representation of the character state and form $\postorderPartial_{\currnode \rate \column}$ on-the-fly such that $\postorderElement_{\currnode \rate \column \state} = 1$ if $ \datum_{ic} = \state$ and 0 for all other states.

The vector-vector multiplications needed for Equation \ref{eq:gradient} are split into two serial operations in lines 7 - 19 and 20 -23.
Lines 7 - 19 perform element-wise multiplications involving $\postorderPartial_{\currnode \rate \column}$ and $\preorderPartial_{\currnode \rate \column}$ serially across rate categories.
Within this section of the algorithm, lines 13 - 16 also perform the matrix-vector multiplication of the infinitesimal rate matrix $\rateMatrix^{(\rate)}$ and $\preorderPartial_{\currnode \rate \column}$.
Taken together, lines 7 - 19 yield $\states$ state-specific entries for the numerator $\phi_{\column}$ and the denominator $\omega_{\column}$.
Lines 20-23 reduce these state-specific entries in parallel to calculate the column-specific contribution to the gradient wrt a BLS parameter $\frac{\partial}{\partial \branchLength_{\currnode}} \log \probability{\data_{\column}}$.

\begin{algorithm}[H]
  \caption{GPU-based parallel computation of the gradient of the log-likelihood wrt all branch-specific parameters}
  \label{alg:gradient}
  \begin{algorithmic}[1]
    \State \textbf{define} COLUMN\_BLOCK\_SIZE (CBS) $=$ number of patterns processed in parallel per thread-block
    \State \textbf{define} PEELING\_BLOCK\_SIZE (PBS) $=$ number of states processed in parallel per inner-loop
    \For {\textbf{all} thread-blocks (node $\currnode=1, \ldots , 2 \numtips - 2$ and column-block $= 1, \ldots ,\lceil C / \text{CBS} \rceil$) \textbf{in parallel}}
      \For{$R$ threads in block (rate $\rate=1, \ldots ,\rates$) \textbf{in parallel}}
          \State \textbf{prefetch} weight $\probability{\gamma_r}$ (reused by all threads in block)
      \EndFor
      \For {
      \textbf{all} threads in block (state $s=1, \ldots ,S$, column $c=1, \ldots ,\text{CBS}$) \textbf{in parallel}}
        \State \textbf{initialize} $\phi_{cs} \gets 0$, $\omega_{cs} \gets 0$
        \For {$\rate= 1,\ldots,\rates$ \textbf{in series}}
          \State \begin{varwidth}[t]{\linewidth-4.5em}
	          \textbf{prefetch} post-order partial likelihood elements $p_{{\currnode}rcs}$ and pre-order partial likelihood  elements $q_{{\currnode}rcs}$ for CBS columns (reused by all threads in block) and \textbf{synchronize}
			\end{varwidth}
          \State \textbf{increment }$\omega_{cs} \gets \omega_{cs} + p_{{\currnode}rcs} \times q_{{\currnode}rcs} \times \probability{\gamma_{r}} $
          \State \textbf{initialize} $\delta \gets 0$
          \For {$\estate = 1,\ldots,\states$ in PBS-sized \textbf{parallel} chunks}
            \State \begin{varwidth}[t]{\linewidth-5.75em}
            \textbf{prefetch} infinitesimal rate elements $Q^{(r)}_{s\estate}$ for PBS states (reused by all threads in the block) and \textbf{synchronize}
            \end{varwidth}
            \State \textbf{increment} $\delta \gets  \delta + Q^{(r)}_{s\estate} \times q_{{\currnode}rc{\estate}}$
          \EndFor
          \State \textbf{increment} $\phi_{cs} \gets \phi_{cs} + p_{{\currnode}rcs} \times \delta \times P(\gamma_{r})$
        \EndFor
      \EndFor
      \For {CBS tasks ($\column = 1,\ldots,\text{CBS}$) with $\states$ threads ($\state = 1,\ldots,\states$) each \textbf{in parallel}}
      	\State \textbf{reduce} $\phi_c \gets \sum_{s=1}^{\states} \phi_{cs}$ and $\omega_c \gets \sum_{\state=1}^{\states} \omega_{cs}$
      	\State \textbf{return}
		$\phi_c / \omega_c$
      	as column-specific contribution
      	$\frac{\partial}{\partial \branchLength_{\currnode}} \log \probability{\data_{\column}}$
      \EndFor
    \EndFor
  \end{algorithmic}
\end{algorithm}

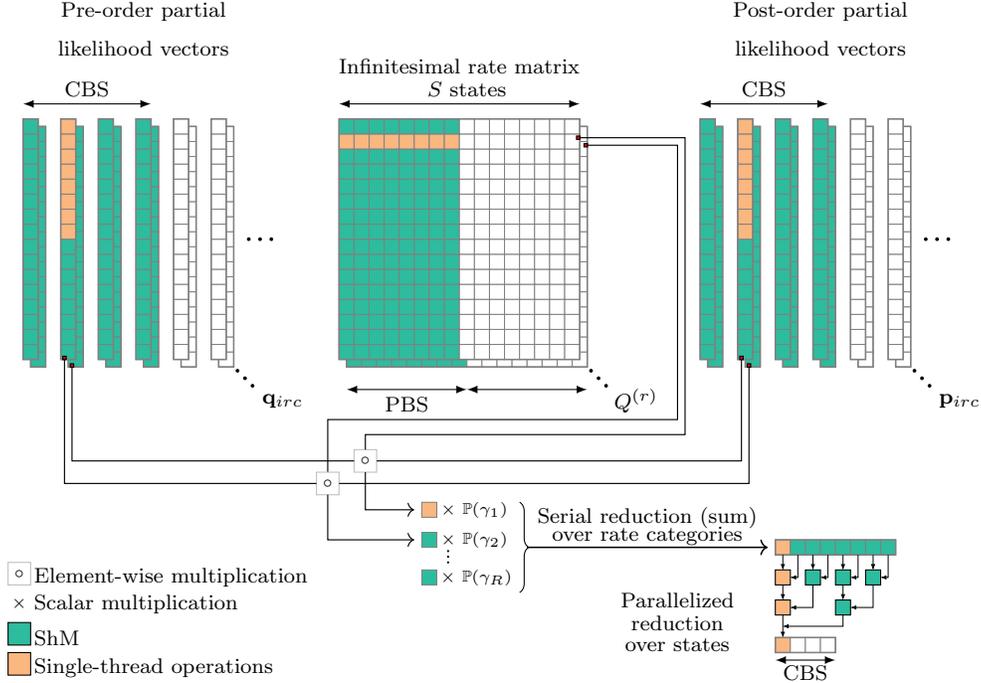
\begin{figure}[H]
  \centering
  \begin{tikzpicture}
    \newcommand\cellwidth{0.2}

    \newcommand\px{0}
    \newcommand\pyone{0}
    \node [draw=none, align=center] at (\px+1.8,\pyone+4.8) {\scriptsize{Pre-order partial}\\\scriptsize{likelihood vectors}};

    \draw [latex-latex] (\px + 0.2, \pyone+3.8) -- (\px+1.9, \pyone + 3.8) node [black,midway, below] {};
    \node [] at (\px+1.05, \pyone+4) {\scriptsize{CBS}};

    \dotplaceholder{\px+2.95}{\pyone+0.35}{$\preorderPartial_{\currnode \rate \column}$};

    \gridpartials{\px + 0.1}{\pyone+ 0.1}{6}{16}{\cellwidth}{-1}{-1}{4}{-1}
    \gridpartials{\px}{\pyone+ 0.2}{6}{16}{\cellwidth}{2}{8}{4}{-1}

    \foreach \x in {0.1,0.25,0.4}{
    	\filldraw [black] (\px+3.1+\x,\pyone+2) circle (0.5pt);
    }

    \newcommand\Qx{4.2}
    \newcommand\Qy{0}
    \node [draw=none, align=center] at (\Qx+1.8,\Qy+4.3) {\scriptsize{Infinitesimal rate matrix}};
    \node at (\Qx + 0.3 + 1.6,\Qy+4) {\scriptsize{$S~\text{states}$}};
    \draw [latex-latex] (\Qx + 0.2,\Qy+3.8) -- (\Qx + 0.2 + 3.2,\Qy+3.8);

    \grid{\Qx + 0.1}{\Qy+0.1}{16}{16}{\cellwidth}{-1}{15}{8}{17}
    \grid{\Qx}{\Qy+ 0.2}{16}{16}{\cellwidth}{-1}{15}{8}{17}

	\node at (\Qx + 0.3 + 0.8,  \Qy - 0.2) {\scriptsize{PBS}};
	\draw [latex-latex] (\Qx + 0.3, \Qy) -- (\Qx+0.3 + 1.6, \Qy);
	\draw [latex-latex] (\Qx + 0.3 + 1.6, \Qy) -- (\Qx+0.3 + 3.2, \Qy );

    \dotplaceholder{\Qx+3.45}{\Qy+0.35}{$Q^{(\rate)}$};

    \newcommand\pxtwo{9}
    \newcommand\pytwo{0}
    \node [draw=none, align=center] at (\pxtwo+1.8,\pytwo+4.8) {\scriptsize{Post-order partial}\\\scriptsize{likelihood vectors}};
   \draw [latex-latex] (\pxtwo + 0.2, \pytwo+3.8) -- (\pxtwo+1.9, \pytwo + 3.8) node [black,midway, below] {};
   \node [] at (\pxtwo+1.05, \pytwo+4) {\scriptsize{CBS}};

    \dotplaceholder{\pxtwo+2.95}{\pytwo+0.35}{$\postorderPartial_{\currnode \rate \column}$};
    \gridpartials{\pxtwo + 0.1}{\pytwo+ 0.1}{6}{16}{\cellwidth}{-1}{-1}{4}{-1}
    \gridpartials{\pxtwo}{\pytwo+ 0.2}{6}{16}{\cellwidth}{2}{8}{4}{-1}

	\foreach \x in {0.1,0.25,0.4}{
		\filldraw [black] (\pxtwo+3.1+\x,\pytwo+2) circle (0.5pt);
	}

    \draw [draw=lightgray] (\px, \pyone-2.3) rectangle (\px+0.3, \pyone-2.6) node[pos=0.5] {\tiny{$\circ$}};
    \node [right] at (\px+0.2, \pyone-2.5) {\scriptsize{Element-wise multiplication}};

    \draw [draw=none] (\px, \pyone-2.7) rectangle (\px+0.3, \pyone-3) node[pos=0.5] {\tiny{$\times$}};
    \node [right] at (\px+0.2, \pyone-2.85) {\scriptsize{Scalar multiplication}};

    \draw [fill=SeaGreen] (\px, \pyone-3.1) rectangle (\px+0.3, \pyone-3.4);
    \node [right] at (\px+0.2, \pyone-3.3) {\scriptsize{ShM}};

    \draw [fill=Apricot] (\px, \pyone-3.5) rectangle (\px+0.3, \pyone-3.8);
    \node [right] at (\px+0.2, \pyone-3.7) {\scriptsize{Single-thread operations}};

    \newcommand\wx{\Qx+1.6}
    \newcommand\wy{\Qy-1.5}
    \draw[draw=gray,line width=0.1mm, fill=Apricot] (\wx-0.3, \wy) rectangle (\wx-0.3 + \cellwidth, \wy-\cellwidth);
    \draw[draw=gray,line width=0.1mm, fill=SeaGreen] (\wx-0.3, \wy-0.4) rectangle (\wx-0.3 + \cellwidth, \wy - 0.4 - \cellwidth);
    \draw[draw=gray,line width=0.1mm, fill=SeaGreen] (\wx-0.3, \wy-0.9) rectangle (\wx-0.3 + \cellwidth, \wy - 0.9 - \cellwidth);
    \foreach \x in {0.1,0.175,0.25}{
    	\filldraw [black] (\wx+0.05, \wy-0.9+\x) circle (0.2pt);
    }
    \draw[draw=none] (\wx+\cellwidth *2-0.05, \wy - 1.1) rectangle (\wx+\cellwidth *2+0.1, \wy - 1.1 + \cellwidth) node[pos=0.5] {\tiny{$\times$} $\probability{\gamma_{\rates}}$};
    \draw [decorate,decoration={brace,amplitude=3.5pt}] (\wx+1, \wy) -- (\wx+1, \wy-1.2) node [black,below,midway] {};

    \foreach \offsetx/\offsety/\offsetextra [count=\offseti] in {0/0/0, 0.1/0.1/0.1} {
    	\tikzmath{
    		\oponex = \Qx+0.55-\offsetx-4*\offsetextra;
    		\oponey = \wy+0.55-2*\offsety-\offsetextra;
    		\opwidth=0.15;
 			\tmxone = \Qx + 4.7 + \offsetx;
    		\tmyone = \Qy + 3.25 + \offsety;
    		\tmxtwo = \Qx + 0.05 + \offsetx +4* \offsetextra;
    		\tmytwo = \wy+0.6 - \offsety - \offsetextra;
    		\pvxone = \px-\offsetx + 0.85;
    		\pvyone = \pyone+\offsety + 0.35;
    		\pvxtwo = \Qx + 0.4 - \offsetx -4* \offsetextra;
    		\pvytwo = \wy+0.55-2*\offsety-\offsetextra;
    		\pvxthree = \pxtwo+\offsetx + 0.75;
    		\pvythree = \pytwo-\offsety +0.45;
    		\pvxfour = \Qx + 0.7 - \offsetx-4*\offsetextra;
    		\pvyfour = \wy+0.55-2*\offsety-\offsetextra;
    		\outxone = \Qx+0.55-\offsetx-4*\offsetextra;
    		\outyone = \wy+0.55-2*\offsety-\offsetextra - 0.15;
    		\outxtwo = \wx-0.4;
    		\outytwo = \wy-0.1-4*\offsetextra;
    	}

    	\draw [draw=lightgray] (\oponex-\opwidth, \oponey-\opwidth) rectangle (\oponex+\opwidth, \oponey+\opwidth) node[pos=0.5] {\tiny{$\circ$}};

     	\draw[{Square[width=0.6mm,length=0.6mm,fill=red]}-,line width=0.15mm]  (\tmxone - 1.25 - 2 * \offsetx, \tmyone) -- (\tmxone, \tmyone)  -- (\tmxone, \tmytwo + 0.5) -- (\tmxtwo, \tmytwo+0.5) --  (\tmxtwo, \tmytwo  - 0.2+ 5 * \offsety);

		\draw[{Square[width=0.6mm,length=0.6mm,fill=red]}-,line width=0.15mm] (\pvxone, \pvyone) -- (\pvxone, \pvytwo) -- (\pvxtwo, \pvytwo);

      	\draw[{Square[width=0.6mm,length=0.6mm,fill=red]}-,line width=0.15mm] (\pvxthree, \pvythree) -- (\pvxthree , \pvyfour) -- (\pvxfour, \pvyfour);

      	\draw[->,line width=0.15mm] (\outxone, \outyone) -- (\outxone,\outytwo) -- (\outxtwo, \outytwo);

      	\draw [draw=none] (\wx+\cellwidth *2-0.15,\wy-0.1-4*\offsetextra+0.15) rectangle (\wx+\cellwidth *2+0.15,\wy-0.1-4*\offsetextra-0.15) node[pos=0.5] {\tiny{$\times$} $\probability{\gamma_{\offseti}}$};
    }

    \node [] at (\wx+2.7, \wy-0.2) {\scriptsize{Serial reduction (sum)}};
    \node [] at (\wx+2.7, \wy-0.45) {\scriptsize{over rate categories}};
    \draw[->,line width=0.2mm] (\wx+1.1,\wy-0.6) -- (\wx+4.3,\wy-0.6);

    \newcommand\prx{\wx+4.2}
    \newcommand\pry{\wy-0.9}
    \grid{\prx}{\pry}{8}{1}{\cellwidth}{1}{2}{8}{2}
    
    \foreach \layer [count=\layeri] in {0,1,2,3} {
    	\tikzmath{
    		\layerx = \prx+0.3+0.4*\layer;
    		\layery = \pry+0.2;
    	}
    	\draw [-{Latex[scale=0.5]}, line width = 0.1mm] (\layerx,\layery) -- (\layerx,\layery-0.2);
   		\draw [-{Latex[scale=0.5]}, line width = 0.1mm] (\layerx + \cellwidth,\layery) -- (\layerx + \cellwidth,\layery-0.3) -- (\layerx + 0.1, \layery-0.3);
   		\ifthenelse{\layer=0} 
   		{
   			\draw [fill=Apricot] (\layerx-0.1,\layery-0.2) rectangle (\layerx+0.1,\layery-0.4);
   		}
   		{
   			\draw [fill=SeaGreen] (\layerx-0.1,\layery-0.2) rectangle (\layerx+0.1,\layery-0.4);
   		}	
    }
	
	\foreach \layer [count=\layeri] in {0,1} {
		\tikzmath{
			\layerx = \prx+0.3+0.8*\layer;
			\layery = \pry-0.2;
		}
		\draw [-{Latex[scale=0.5]}, line width = 0.1mm] (\layerx,\layery) -- (\layerx,\layery-0.2);
		\draw [-{Latex[scale=0.5]}, line width = 0.1mm] (\layerx + \cellwidth*2,\layery) -- (\layerx + \cellwidth*2,\layery-0.3) -- (\layerx + 0.1, \layery-0.3);
		\ifthenelse{\layer=0} 
		{
			\draw [fill=Apricot] (\layerx-0.1,\layery-0.2) rectangle (\layerx+0.1,\layery-0.4);
		}
		{
			\draw [fill=SeaGreen] (\layerx-0.1,\layery-0.2) rectangle (\layerx+0.1,\layery-0.4);
		}
	}

	\tikzmath{
		\layerx = \prx+0.3;
		\layery = \pry-0.6;
	}
	\draw [-{Latex[scale=0.5]}, line width = 0.1mm] (\layerx,\layery) -- (\layerx,\layery-0.3);
	\draw [-{Latex[scale=0.5]}, line width = 0.1mm] (\layerx + \cellwidth*4,\layery) -- (\layerx + \cellwidth*4,\layery-0.15) -- (\layerx, \layery-0.15);

    \node [] at (\wx+3.1, \wy-1.3) {\scriptsize{Parallelized}};
    \node [] at (\wx+3.1, \wy-1.6) {\scriptsize{reduction}};
    \node [] at (\wx+3.1, \wy-1.9) {\scriptsize{over states}};

	\grid{\prx}{\pry-1.3}{4}{1}{\cellwidth}{1}{2}{1}{2}
	
	\node at (\prx+0.6,  \pry - 1.375) {\scriptsize{CBS}};
	\draw [latex-latex] (\prx+0.2, \pry-1.2) -- (\prx+1, \pry-1.2);

  \end{tikzpicture}
  \caption{Parallel thread-block design to calculate the column-specific contributions to the gradient of the log-likelihood wrt all BLS parameters for all columns $\columns$. One block evaluates the column-specific contribution to the gradient wrt a BLS parameter $\frac{\partial}{\partial \branchLength_{\currnode}} \log \probability{\data_{\column}}$ for CBS columns by prefetching $\states \times \text{CBS}$ entries from the pre- and post-order partial likelihood vectors ${\preorderPartial}_{\currnode \rate \column}$ and $\postorderPartial_{\currnode \rate \column}$ in parallel (Algorithm \ref{alg:gradient}: Line 10) and $\states \times \text{PBS}$ entries at a time from $\rateMatrix^{(\rate)}$ within an inner serial loop (Algorithm \ref{alg:gradient}: Lines 13 - 16). Each block performs a serial reduction over rate categories (Algorithm \ref{alg:gradient}: Lines 7 - 19) and a parallelized reduction over states (Algorithms \ref{alg:gradient}: Lines 20 - 23). Entries fetched from global memory (GM) into shared memory (ShM) are indicated in green and an instance of a single-thread operation is shown in orange.}
  \label{fig:gradientKernel}
\end{figure}

The CPU implementation  of Equation \ref{eq:gradient} only parallelizes calculations across conditionally independent blocks of the sequence alignment but Algorithm \ref{alg:gradient} takes advantage of the GPU to introduce a great deal of fine-scale parallelization across nodes and columns.
Similar to Algorithm \ref{alg:preorderPartial}, most of the work per thread involves effectively caching values in ShM to be reused by the threads within a thread-block.
Figure \ref{fig:gradientKernel} outlines the parallel thread-block design for Algorithm \ref{alg:gradient}.
Since the weights $\probability{\siteSpecificRate_r}$ of the rate categories are the same for all columns, we use $\rates$ threads within a block to fetch these weights into ShM.
Equation \ref{eq:gradient} shows that calculating $\frac{\partial}{\partial \branchLength_{\currnode}} \log \probability{\data_{\column}}$ for node $\currnode$ and column $\column$ involves taking a weighted sum of rate-specific entries.
Each rate-specific entry depends on the same $\states$ partial likelihoods from each of the partial likelihood vectors ${\preorderPartial}_{\currnode \rate \column}$ and $\postorderPartial_{\currnode \rate \column}$.
Based on this observation, we construct $(2\numtips -2) \times  \lceil C/CBS \rceil$ thread-blocks with each thread-block sharing $\states \times \text{CBS}$ threads.
All $\states \times \text{CBS}$ threads within a block concurrently fetch $\states$-lengthed vectors $\preorderPartial_{\currnode \rate \column}$ and $\postorderPartial_{\currnode \rate \column}$ for CBS columns from GM into ShM.
We also observe that the rate-specific entries for all columns depend on the same infinitesimal rate matrix $\rateMatrix^{(\rate)}$.
This allows us to use $\states$ threads within a block to fetch columns from the rate matrix in PBS chunks  from GM into ShM as described previously in Algorithm \ref{alg:preorderPartial}.
Each thread-block computes $\frac{\partial}{\partial \branchLength_{\currnode}} \log \probability{\data_{\column}}$ for one node and CBS columns.

Since columns in the sequence alignment are assumed to be independent, arising from conditionally independent CTMCs acting along each branch, we can calculate $\frac{\partial}{\partial \branchLength_{\currnode}} \log \probability{\data}$ by reducing the partial derivatives across all columns $\columns$ according to Equation \ref{eq:gradientSumAcrossColumns}.
We wrote a `multipleNodeSiteReduction' kernel to perform this reduction in parallel with each thread-block reading in 128 column-specific contributions to the gradient.
For certain models such as those that assume a strict molecular clock, it might more be convenient to further reduce the partial derivatives across a set of branches and report a single value.
We currently perform this final reduction on the CPU to maintain the ability to specify any desired set of branches.
The design constants CBS and PBS for Algorithm \ref{alg:gradient} are the same as Algorithm \ref{alg:preorderPartial} with $\text{CBS}=16$ for nucleotide models and $\text{CBS} = 8$ and $\text{PBS}=8$ for codon models.

Our current model assumes $Q^{(r)} = \siteSpecificRate_{r} \rateMatrix$, but we have left the algorithm for the more general case when $Q^{(r)}$ can vary arbitrarily across rate categories.
Without this generalization, we could reduce some memory transactions by reading $\siteSpecificRate_{1}, \ldots, \siteSpecificRate_{\rates}$ and $\rateMatrix$ once and forming $Q^{(r)}  = \siteSpecificRate_{\rate} \rateMatrix$ on-the-fly.
Another limitation of Algorithm \ref{alg:gradient} arises when $R > S \times \text{CBS}$ even though $\rates \leq 10$ is adequate for most common phylogenetic analyses \citep{jiaImpactModellingRate2014}.
In such a case, lines 7 - 19 in Algorithm \ref{alg:gradient} can be executed serially on chunks of rate categories that can be controlled by introducing a new, rate-block-size design constant.

\section{Results}

We use three datasets to illustrate the performance gains afforded by the algorithms presented in this paper: (1) a dengue virus dataset of 997 genomes with 6,869 unique nucleotide site patterns across 10 genes and 3,343 unique site patterns when translated into a 61 state universal codon model, (2) a carnivores dataset of 62 genomes with 5,565 unique nucleotide site patterns and 3,600 unique site patterns when translated into a 61 state vertebrate mitochondrial codon model and (3) a yeast dataset of 49 genomes with 12,878 unique nucleotide site patterns and 22,151 unique site patterns when translated into a 61 state universal codon model. 
To provide comprehensive estimates of performance gains, we use three systems with varied technical specifications.
System 1 is equipped with a 10-core 3.3 GHz Intel Xeon W-2155 processor, 32 GB 2.6 GHz DDR4 RAM and an NVIDIA Quadro GV100 GPU with 5,120 cores running at 1.1 GHz and 32 GB global memory.
System 2 is equipped with a 20-core 2.2 GHz Intel Xeon E5-2698 processor, 512 GB 2.6 GHz DDR4 RAM and an NVIDIA Tesla V100 GPU with 10,240 cores running at 1.53 GHz and 32 GB global memory.
System 3 is equipped with a 48-core 2.3 GHz AMD EPYC 7642 Processor, 512 GB DDR4 RAM and an AMD MI50 GPU with 3840 cores running at 1.73 GHz and 32 GB global memory.

\begin{figure}[H]
  \center
  \includegraphics[width=\textwidth]{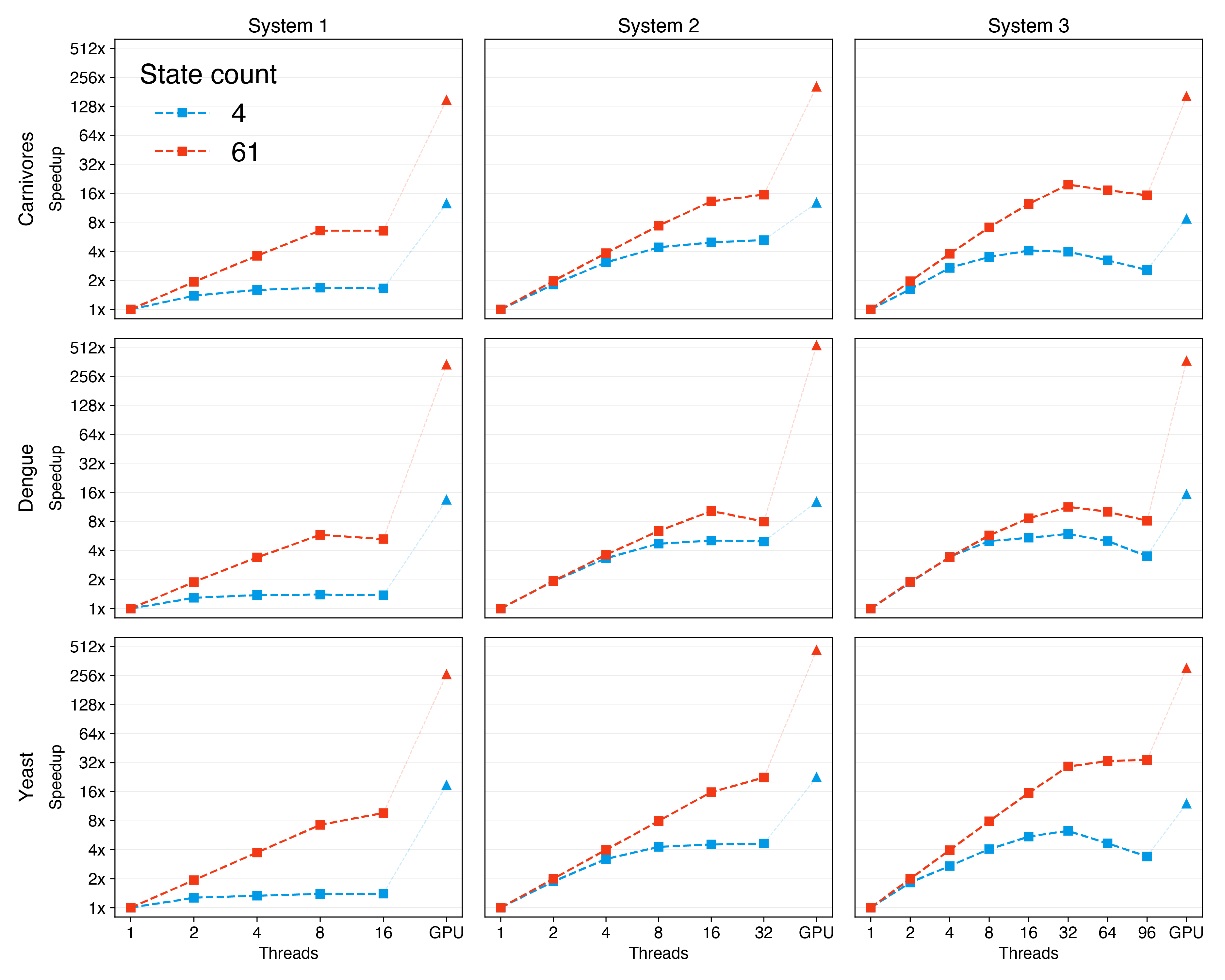}
  \caption{Speedup of GPU and multi-core CPU instances over a single CPU thread for five MCMC iterations to infer branch-specific evolutionary rates for a GTR model with state-space size of 4 and a Yang codon model with a state-space size of 61. The comparisons are reported for three datasets on three different systems. Speedup factors are on a log-scale.}
  \label{figure:speedup}
\end{figure}

For each dataset, we infer branch-specific evolutionary rates given fixed trees from a Bayesian analysis for two substitution models, the general time reversible (GTR) nucleotide model \citep{lanaveNewMethodCalculating1984} including discrete gamma-distributed rate variation with four categories and the Yang codon model \citep{yangCodonSubstitutionModelsHeterogeneous2000}.
We perform this analysis on each of the three systems and measure the wall-time of five iterations of the MCMC using HMC as described in \citet{alexrrw} and implemented in the Bayesian phylogenetic reconstruction software package BEAST \citep{Suchard2018beast}.
We report the corresponding speedup of multi-threaded CPU and GPU instances over a single-threaded CPU in Figure \ref{figure:speedup}.
For the GTR model, we see that performance on the CPU reaches saturation at 32 threads on systems 2 and 3 with a near 4-fold speedup as compared to single-core performance on all three datasets.
On system 1 which is equipped with a less powerful CPU, we see that the multi-threaded CPU implementation offers more modest speedups of less than 2-fold.
On system 3 which has a CPU with 48 cores, we see that increasing the number of threads over 32 to 64 and 96 results in longer wall-times.
Our algorithms that utilize the GPU offer a higher speedup of near 16-fold on all three systems.

We consistently see higher performance gains for the Yang codon model with a state-space  size of 61 compared to the 4-state GTR model on both the CPU and GPU\@.
For the Yang codon model, the performance of the CPU implementation reaches saturation at 8, 16, and 32 threads on systems 1, 2 and, 3  respectively.
On the CPU, we see a maximum speedup of near 8-fold on system 1 and near 16-fold on systems 2 and 3 for all three datasets with the exception of the yeast dataset on system 3.
The yeast dataset yields a much higher number of unique site patterns when translated into a 61 state codon model compared to the other two datasets and the powerful 48-core CPU on system 3 offers a maximum speedup of near 32-fold.
As we previously observed for the GTR model, using more than 32 CPU threads on system 3 results in a decrease in performance.
The GPU for the Yang codon model offers speedups of greater than 128-fold on all three systems that far exceeds the performance gain that can be obtained from a multi-threaded CPU.

To determine whether increasing the state-space size beyond 61 would result in higher performance gains on the GPU, we also inferred branch-specific evolutionary rates for the yeast dataset using a Markov-modulated model (MMM) composed of two Yang codon models yielding a combined state-space size of 122.
The speedups for the MMM model are very similar to the Yang codon model, showing that the resources on the GPU are maximally utilized at a state-space size of 61 (Figure \ref{figure:statespace61vs122}).

In addition to the wall-time, we also measured the time spent in calculating the pre-order partial likelihood vectors and the gradient of the log-likelihood wrt all BLS parameters for the 4-state GTR model.
We report the speedup of our GPU implementation relative to a single-threaded CPU in Table \ref{table:functionSpeedUp}.
We see that the GPU implementation offers over 19-fold improvement for calculating the pre-order partial likelihood vectors and over 7-fold improvement for calculating the gradient across all three systems.

Since our algorithm for the pre-order traversal builds on the post-order traversal algorithm described in \citet{suchardManycoreAlgorithmsStatistical2009}, we measured the time spent in each traversal while inferring branch-specific evolutionary rates under the Yang codon model using the carnivores dataset (Table \ref{table:gpuTimingsCodon}).
We observe that the average time per function call for the post-order traversal (126,096 ns) is faster than the pre-order traversal (177,096 ns).
The pre-order kernel is also called nearly two times more than the post-order kernel since pre-order partial likelihoods must be calculated at the tip nodes, whereas the $\datum_{ic}$ is directly observed for the post-order partial likelihoods at the tip nodes.
This latter point also explains why the small speed-difference is unsurprising.
In spite of the reduced burden on ShM, the pre-order traversal requires larger GM transactions than the post-order traversal immediately above and at the tip nodes; when $\datum_{ic}$ is observed, the post-order partial likelihoods are sparse, while the pre-order partial likelihoods are always dense.
Overall, the pre-order kernel takes 44.62\% of the total wall-time while the post-order kernel takes 17.21\%.
In addition to Algorithms \ref{alg:preorderPartial} and \ref{alg:gradient}, we also implemented two additional kernels, namely, the matrixTranspose kernel to transpose transition matrices for models with large state-space sizes and the nodeSiteReduction kernel to reduce the column-specific contributions to the gradient across columns.
Both these kernels require very little execution time, with the matrixTranspose kernel taking roughly 0.26\% and the nodeSiteReduction kernel taking roughly 0.04\% of the total wall-time (Table \ref{table:gpuTimingsCodon}).

\begin{table}[H]
  \centering
  \begin{tabular}{lrrrrrrr}
		\toprule
		\multirow{2}{*}{Dataset} & \multicolumn{4}{c}{Preorder Traversal} & \multicolumn{3}{c}{Gradient calculation} \\
		\cline{2-4}
		\cline{6-8}
		& System 1 & System 2 & System 3 & & System 1 & System 2 & System 3\\
		\midrule
		Carnivores & 23 & 27 & 19 & & 23 & 32 & 16\\
		Dengue & 29 & 31 & 24 & & 7 & 8 & 14 \\
		Yeast & 31 & 39 & 24 & & 19 & 35 & 18 \\
		\bottomrule
  \end{tabular}
  \caption{Speedup in calculating pre-order partial likelihood vectors and the gradient of the log-likelihood for a GTR model on the GPU relative to a single-threaded CPU.}
  \label{table:functionSpeedUp}
\end{table}

\begin{figure}[H]
	\center
	\includegraphics[width=0.75\textwidth]{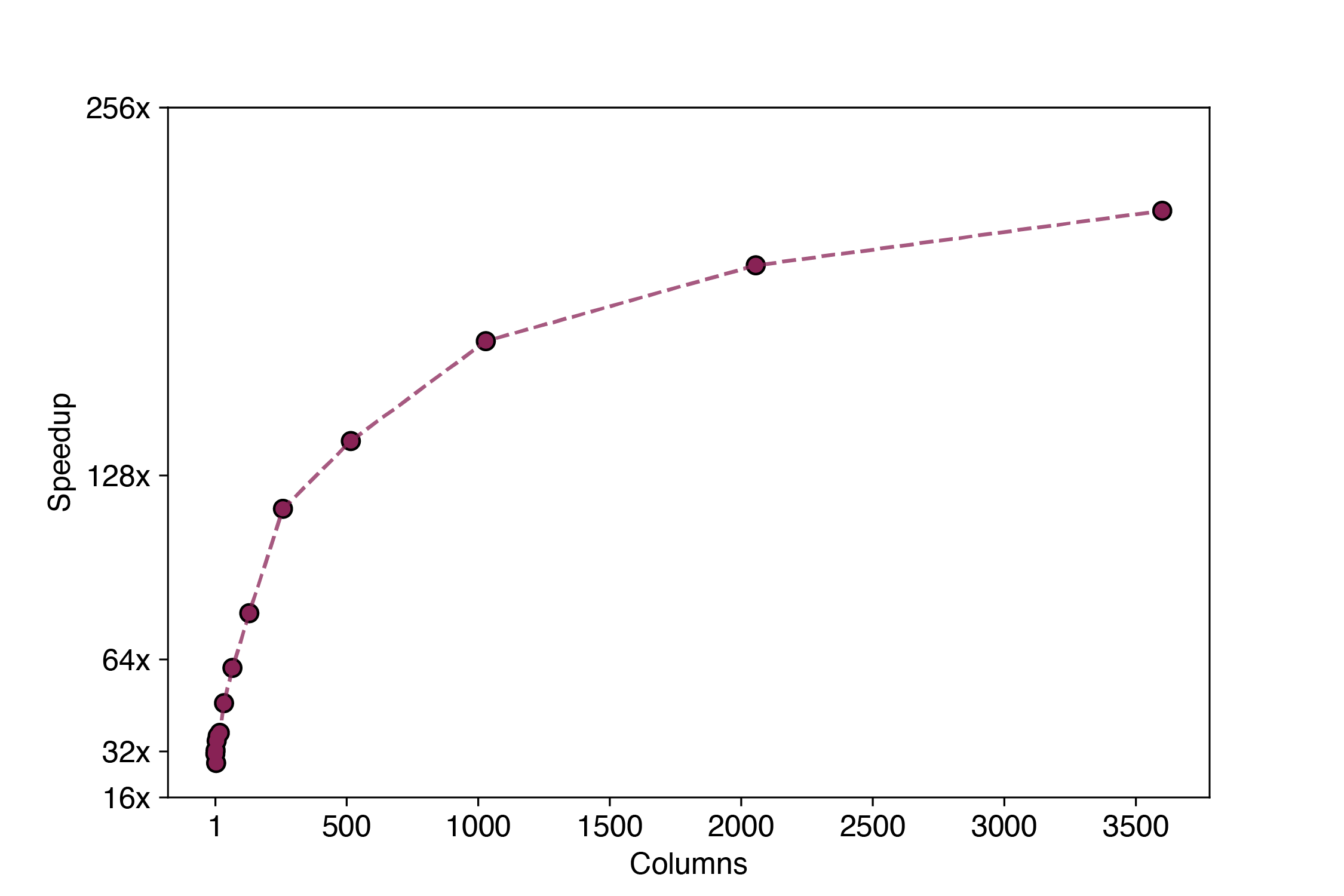}
	\caption{Speedup on the GPU relative to a single threaded CPU scaled by the number of unique alignment columns $\columns$ using the codon-alignment of the carnivores dataset. Speedup factors are on the log-scale.}
	\label{figure:patternsspeedup}
\end{figure}

Apart from state-space size $\states$, another dimension that influences the effectiveness of parallelization on GPUs is the number of unique site patterns or alignment columns $\columns$.
For a small number of columns, the overhead of caching values in ShM might outweigh the performance gain afforded by the parallelization of the numerical calculations.
As the number of columns increases, we expect performance to increase until it reaches saturation when the resources on the GPU are fully utilized.
To measure how performance scales with the number of columns, we truncated the codon alignment of the carnivores dataset to obtain an increasing number of unique site patterns and report the associated speedup on the GPU over a single-threaded CPU instance in Figure \ref{figure:patternsspeedup}.
Even for a single column, we observe a speedup of nearly 31-fold with performance reaching saturation at $\columns = 1,024$ indicating the maximal utilization of the resources on the GPU at that point followed by marginal increases in performance as more columns are added.

\section{Example}

To demonstrate the utility of the algorithms presented in this paper, we infer the date of the first introduction of West Nile virus into the United States under the Yang codon model with branch-specific evolutionary rates.
West Nile virus is a mosquito-borne RNA virus that was first detected in the United States in New York City in August 1999 \citep{cdcwnv1999}.
Since its first detection on the East Coast in 1999, the virus spread westward across the continental United States and was first detected on the West Coast in California in November 2003 \citep{reisenWestNileVirus2004a}.
The virus has caused over 52,000 cases and over 2,400 deaths as of 2020, making it the leading cause of domestically acquired arbovirus disease in the continental United States \citep{sotoWestNileVirus2022}.

\begin{figure}[H]
  \center
  \includegraphics[height=0.5\textheight]{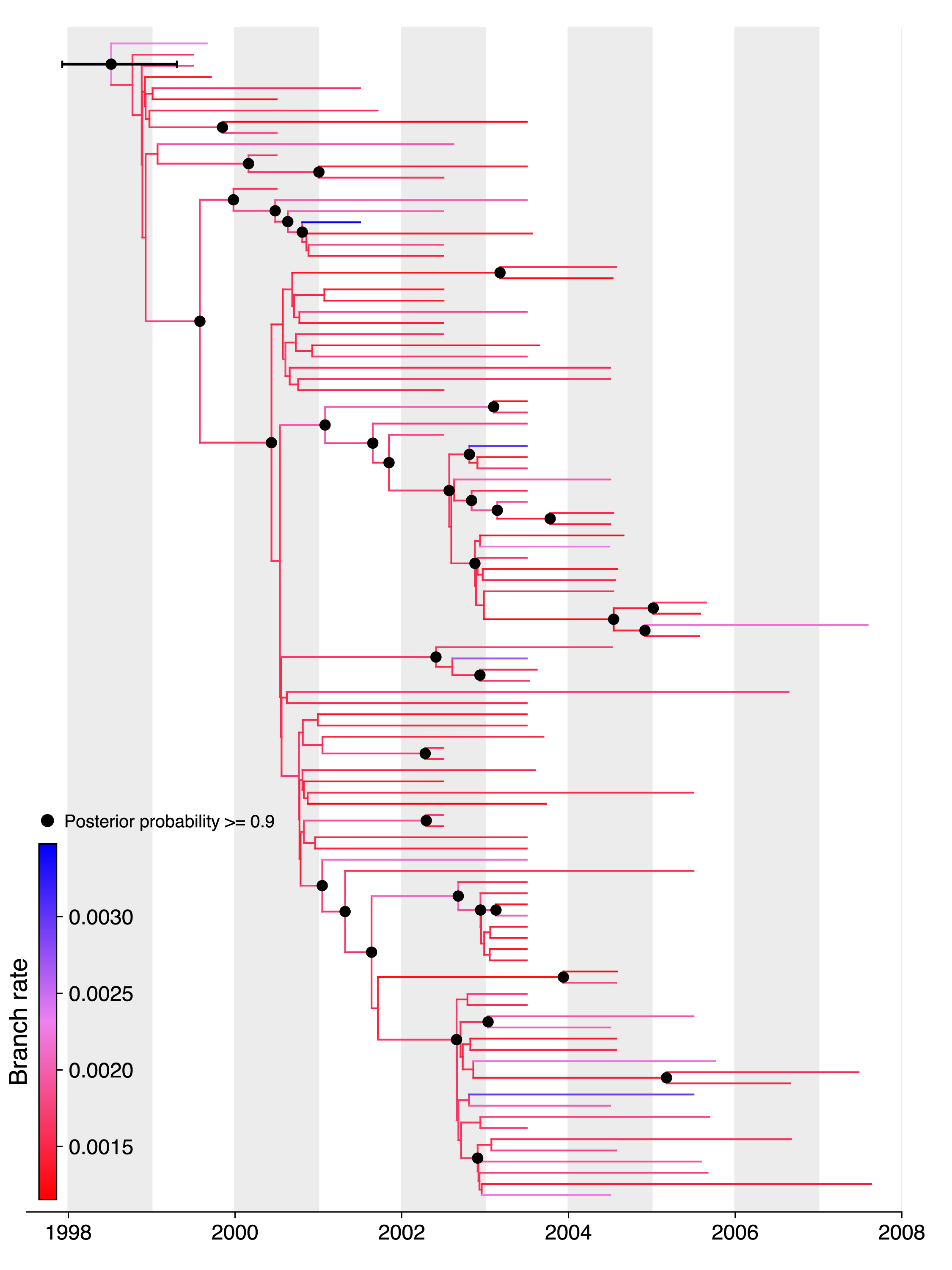}
  \caption{Reconstructed codon-based maximum clade credibility (MCC) tree of 104 West Nile virus genomes sampled in the continental United States between 1999 and 2007.}
  \label{figure:wnvtree}
\end{figure}

Here, we use a dataset of 104 full viral genomes collected in the continental United States between 1999 and 2007 \citep{pybusUnifyingSpatialEpidemiology2012a}.
Each genome encodes for a single polyprotein precursor that is post translationally cleaved into three structural and seven nonstructural proteins \citep{brintonMolecularBiologyWest2002}.
When translated into a 61-state universal codon model, this alignment yields 1,126 unique site patterns.
We infer the age of the root of these genomes from a Bayesian analysis under the Yang codon model with a 4-class discrete-$\Gamma$ model for site rate variation \citep{yangAmongsiteRateVariation1996}, an uncorrelated relaxed clock model \citep{drummondRelaxedPhylogeneticsDating2006} and a skyline non-parametric coalescent prior \citep{drummondSkyline} on the unknown tree.
Figure \ref{figure:wnvtree} displays the maximum clade credibility (MCC) tree inferred using BEAST with HMC over the branch-specific rate scalars, node heights, and the parameters of the skyline population model.
The default transition kernels were used over the remaining random parameters including the unknown tree and codon model parameters.
We ran this analysis for 10 million MCMC iterations with the first 10\% of iterations discarded as burn-in.
The effective sample size (ESS) for all scientifically relevant parameters was above 200.
Convergence of the chain and ESS of parameters were assessed using Tracer \citep{rambautTracer} and the MCC tree was constructed using TreeAnnotator 1.10 and visualized using baltic \citep{evogytisBaltic}.

Table \ref{table:wnv} reports the marginal posterior estimates for the age of the root, the transition:transversion ratio $\kappa$, and the non-synonymous:synonymous rate ratio $\omega$.
The posterior median estimate for the age of the root was 1 August 1998 (95\% highest posterior density [HPD] interval: [September 1997, Feb 1999]) that stands in line with previous studies that inferred the age of the root from a Bayesian analysis under a nucleotide model \citep{pybusUnifyingSpatialEpidemiology2012a}.
This result is also consistent with prior research that suggested a similar introduction time based on circumstantial evidence \citep{lanciottiOriginWestNile1999}.
As reported previously in \citet{anezEvolutionaryDynamicsWest2013}, the virus has been subjected to strong purifying selection since its introduction as evidenced by a posterior median $dN/dS$ ratio $\omega$ estimate of 0.14 (95\% HPD: [0.12, 0.16]).
We performed this analysis on a desktop equipped with a 20-core 2.2 GHz Intel Xeon E5-2698 processor, 512 GB 2.6 GHz DDR4 RAM and an Nvidia Tesla V100 GPU with 10,240 cores running at 1.53 GHz and 32 GB global memory.
Using our algorithms on the GPU, we were able to complete this analysis in just roughly 100 hours, a remarkable improvement when compared to a back of the envelope calculation that estimated this same analysis would take over 1500 days on a CPU.

\begin{table}[H]
  \centering
  \begin{tabular}{lrr}
    \toprule
    Parameter & Posterior median & 95\% Bayesian credible interval \\
    \hline
    Age of root & 1998.58 & (1997.72 - 1999.10) \\
    Transition:transversion rate $\kappa$ & 11.34 & (9.33 - 13.55) \\
    \textit{dN/dS} ratio $\omega$ & 0.14 & (0.12-0.16) \\
    \bottomrule
  \end{tabular}
  \caption{Parameter estimates of Yang codon model for 104 West Nile virus genomes sampled in the continental United States between 1999 and 2007.}
  \label{table:wnv}
\end{table}

\section{Discussion}
The algorithms presented in this paper utilize GPUs to deliver several orders of magnitude speedup to calculate the pre-order partial likelihood vectors and subsequently evaluate the gradient of the data log-likelihood wrt all BLS parameters.
Multi-core CPUs can provide an increase in performance but eventually reach a saturation point beyond which increasing the number of threads results in a decrease in performance.
Our GPU-based algorithms afford performance gains that cannot be achieved trivially by using more threads available on a CPU\@.
The improvement in performance for nucleotide models is noteworthy, but becomes even more remarkable when applied to models with a large state-space size such as codon models.
This performance gain enables Bayesian phylogenetic analyses to infer branch-specific parameters using gradient-based samplers like HMC, as showcased by the example of dating the first introduction of West Nile virus into the continental United States.

Our algorithms also have implications for researchers looking to invest in hardware upgrades in order to increase computational speed in a cost-effective manner.
As of January 2023, AMD Ryzen Threadripper PRO 5975WX with 64 cores (retails for around \$6,500 on Newegg) has the highest number of cores among commercially available CPUs but costs twice as much as an Nvidia Tesla V100 GPU (retails for around \$3,500 on Newegg) which has sufficient double-precision floating point performance for phylogenetic analyses.
Thus, GPUs offer a much lower price-performance ratio compared to CPUs which is of consequence for any decision regarding the purchase of new systems or the upgrade of existing systems for individuals as well as high performance computing clusters at academic and non-academic institutions.
The ubiquity of cloud computing has allowed researchers to perform computationally intensive analyses without having to purchase their own hardware.
Even in this scenario, the cost of a compute instance equipped with a single GPU is more cost-efficient compared to one equipped with a large number of CPU cores.
For instance, on Amazon web services as of January, 2023, the on-demand pricing of p3.2xlarge with 1 Tesla V100 is \$3.06/hr whereas the c6g.8xlarge with 32 threads (vCPUs) costs \$1.088/hr.
However, the speedups afforded by our algorithms using the GPU are $4 - 16$-fold higher than using 32 threads on a CPU making the former more cost-efficient.

There remain several limitations to our current massively parallel algorithms.
Among these, for $\states > 4$, we currently take as input the transpose of a transition probability matrix for the pre-order computation, and we compute this transposition via a separate matrixTranspose kernel.
Currently, we compute all matrix transpositions in parallel to avoid an additional kernel launch for each pre-order evaluation.
This requires storing both the original matrices and their transposes in GM.
In a piece of on-going research, we are utilizing the log-likelihood gradient wrt over 20,000 branch lengths under an $\states = 256$ model.
These transposed matrices require approximately $20,000 \times 256 \times 256 \times 8 \, \text{bytes} \approx 10\text{GB}$ of additional RAM in double-precision, severely restricting the range of GPUs that we can employ.
Memory constraints due to too many columns can be addressed by using multiple GPUs but this is not a solution for transition matrix memory constraints.
We could, however, split this operation across multiple CUDA streams with each stream concurrently computing the transpose of a single transition matrix and the corresponding pre-order partial likelihoods.
This would remove the need to store all the transition matrices and their transposes in GM simultaneously.
There are also ways to improve performance on the CPU to evaluate the log-likelihood gradient.
The current CPU implementation to evaluate the gradient only allows concurrent execution across conditionally independent blocks of columns in the alignment but additional parallelization across nodes can also be exploited.
However, its unlikely that any additional parallelization on the CPU would lead to a speedup that is on the same scale as the improvement achieved by using GPUs.

While the algorithms in this study were presented in a Bayesian framework, they also have applications in non-linear optimization in a maximum-likelihood framework.
To make our algorithms available to the broader audience of developers working on statistical phylogenetics, we provide implementations in the open-source BEAGLE v4.0.0 library \citep{ayresBEAGLEImprovedPerformance2019} that uses OpenMP for multi-core CPUs, CUDA and OpenCL for GPUs.

\section*{Data Availability}

Complete BEAST XML files and associated scripts to reproduce the three performance study datasets and WNV example are available at \url{https://github.com/suchard-group/parallel_gradients_supplement}. The log files from the benchmarking and the results from the WNV analysis have been deposited at \url{https://doi.org/10.5281/zenodo.7697474}.

\section*{Software Availability}

The algorithms described in this paper have been implemented in BEAGLE v4.0.0 available at \url{https://github.com/beagle-dev/beagle-lib/releases/tag/v4.0.0}. 

\section*{Acknowledgments}

This work was supported through National Institutes of Health grants R01 AI153044 and R01 AI162611.
We gratefully acknowledge support from NVIDIA Corporation and Advanced Micro Devices, Inc.~with the donation of parallel computing resources used for this research.
PL and MAS acknowledge support from the European Union's Horizon 2020 research and innovation programme (grant agreement no. 725422-ReservoirDOCS) and from the Wellcome Trust through project 206298/Z/17/Z.
PL acknowledges support from the Research Foundation - Flanders (‘Fonds voor Wetenschappelijk Onderzoek - Vlaanderen’, G0D5117N and G051322N) and from the European Union's Horizon 2020 project MOOD (grant agreement no. 874850).
GB acknowledges support from the Research Foundation - Flanders (‘Fonds voor Wetenschappelijk Onderzoek - Vlaanderen’, G0E1420N and G098321N) and from the Internal Funds KU Leuven under grant agreement C14/18/094.
FAM is an Investigator of the Howard Hughes Medical Institute.

\newcommand{\beginsupplement}{%
	\setcounter{table}{0}
	\renewcommand{\thetable}{S\arabic{table}}%
	\setcounter{figure}{0}
	\renewcommand{\thefigure}{S\arabic{figure}}%
}

\bibliographystyle{biometrika}
\bibliography{parallel_gradients}

\clearpage

\beginsupplement
\section{Supplementary Material}

\begin{table}[H]
	\centering
	\sisetup{table-auto-round}
	\begin{tabular}{
		l
		S[table-format=4]
		S[table-format=4]
		S[table-format=1.2]
	}
		\toprule
		{Name} & {Number of calls} & {Time per function call (ns)} & {Percentage (\%)} \\
		\midrule
			preOrderPartials &   1464 &     177096 &   44.620782 \\
			gradient &     24 &    7515810 &   31.043695 \\
			postOrderPartials &    793 &     126096 &   17.209172 \\
			matrixTranspose &     12 &     126000 &    0.260219 \\
			nodeSiteReduction &     12 &      19027 &    0.039294 \\
			Other kernels &    846 &      46888 &    6.826837 \\
		\bottomrule
	\end{tabular}
	\caption{Execution time of individual kernels in a single MCMC iteration to infer the branch specific evolutionary rates using a Yang codon model for the carnivores dataset. Time is reported in nanoseconds (ns).}
	\label{table:gpuTimingsCodon}
\end{table}

\begin{figure}[H]
	\centering
	\newcommand\Qx{0}
	\newcommand\Qy{3}
	\newcommand\cellwidth{0.2}
	\newcommand\cellwidthtwo{0.14}
	\begin{tikzpicture}
		\node at (\Qx + 0.3 + 1.6,\Qy+4.3) {\scriptsize{Transition probability matrix}};
		\node at (\Qx + 0.3 + 1.6,\Qy+3.9) {\scriptsize{$S~\text{states}$}};
		\draw [latex-latex] (\Qx + 0.3,\Qy+3.7) -- (\Qx + 0.3 + 3.2,\Qy+3.7);

		\grid{\Qx + 0.2}{\Qy}{16}{16}{\cellwidth}{-1}{15}{8}{17}
		\grid{\Qx + 0.1}{\Qy+ 0.1}{16}{16}{\cellwidth}{-1}{15}{8}{17}

		\node at (\Qx + 0.4 + 0.8,  \Qy - 0.2) {\scriptsize{PBS}};
		\draw [latex-latex] (\Qx + 0.4, \Qy) -- (\Qx+0.4 + 1.6, \Qy);
		\draw [latex-latex] (\Qx + 0.4 + 1.6, \Qy) -- (\Qx+0.4 + 3.2, \Qy );

		\tikzmath{
			\tmlegendxone  = \Qx;
			\tmlegendyone = \Qy+0.4;
			\tmlegendxtwo = \Qx;
			\tmlegendytwo = \Qy;
		}

		\node [left] at (\tmlegendxone, \tmlegendyone) {\scriptsize{
				$\probabilityMatrix{r}{\branchLength_{\sibling}}$
		}};

		\node [left] at (\tmlegendxtwo, \tmlegendytwo) {\scriptsize{
				$\probabilityMatrix{r}{\branchLength_{\currnode}}$
		}};

		\draw[{Circle[width=0.6mm,length=0.6mm]}-,line width=0.15mm] (\tmlegendxone+0.4, \tmlegendyone) -- (\tmlegendxone-0.1, \tmlegendyone);
		\draw[{Circle[width=0.6mm,length=0.6mm]}-,line width=0.15mm] (\tmlegendxtwo+0.5, \tmlegendytwo+0.25) -- (\tmlegendxtwo+0.1, \tmlegendytwo+0.25) -- (\tmlegendxtwo+0.1, \tmlegendytwo) -- (\tmlegendxtwo-0.1, \tmlegendytwo);

		\newcommand\px{5.5}
		\newcommand\py{3.1}

    	\gridpartials{\px+0.1}{\py-0.1}{8}{16}{\cellwidth}{-1}{-1}{6}{-1}
		\gridpartials{\px}{\py}{8}{16}{\cellwidth}{2}{8}{6}{-1}

		\tikzmath{
			\pvlegendxone  =\px+4.15;
			\pvlegendyone = \tmlegendyone;
			\pvlegendxtwo = \px+4.15;
			\pvlegendytwo = \tmlegendytwo;
		}

		\node [right] at (\pvlegendxone, \pvlegendyone) {\scriptsize{
				$\postorderPartial_{\sibling \rate \column}$

		}};
		\node [right] at (\pvlegendxtwo, \pvlegendytwo) {\scriptsize{
				$\postorderPartial_{\currnode \rate \column}$
		}};

		\draw[{Circle[width=0.6mm,length=0.6mm]}-,line width=0.15mm] (\pvlegendxone - 0.4, \pvlegendyone) -- (\pvlegendxone+0.1, \pvlegendyone);
		\draw[{Circle[width=0.6mm,length=0.6mm]}-,line width=0.15mm] (\pvlegendxtwo-0.3, \pvlegendytwo + 0.275) -- (\pvlegendxtwo-0.3, \pvlegendytwo) -- (\pvlegendxtwo+0.1, \pvlegendytwo);

		\node [right] at (\px,\py+4.2) {\scriptsize{Partial likelihood vectors}};

		\draw [latex-latex] (\px + 0.2, \py+3.6) -- (\px+2.9, \py + 3.6) node [black,midway, below] {};
		\node [] at (\px+1.5, \py+3.8) {\scriptsize{CBS}};

		\node[left] at (\px+0.2,\py+2) {\scriptsize{$\states~\text{states }$}};
		\draw [latex-latex] (\px,\py+0.2) -- (\px,\py+3.4);

		\foreach \x in {0.1,0.25,0.4}{
			\filldraw [black] (\px+4.1+\x,\py+1.8) circle (0.5pt);
		}

		\newcommand\ox{0.25}
		\newcommand\oy{1}


		\grid{\ox}{\oy}{64}{1}{\cellwidthtwo}{2}{-1}{-1}{2}
		\grid{\ox-0.1}{\oy-0.1}{64}{1}{\cellwidthtwo}{2}{-1}{-1}{2}

		\draw [decorate,decoration={brace,amplitude=3.5pt}]
		(\ox+9.1, \oy-0.1) -- (\ox+0.15, \oy-0.1) node [black,midway] {};
		\node [] at (\ox+4.65, \oy-0.4) {\scriptsize{$S \times $\text{CBS} entries $=$ CBS vectors}};

		\newcommand\wx{\Qx+1.6}
		\newcommand\wy{\Qy-1}

		\foreach \offsetx/\offsety/\offsetextra [count=\offseti] in {0/0/0, 0.15/0.1/0.3} {

			\tikzmath{
				\oponex = \Qx+4.5+\offsetx+\offsetextra;
				\oponey = \wy+0.55-\offsety-\offsetextra;
				\opwidth = 0.15;
				\tmxone = \Qx + 0.55 + 3.35- \offsetx;
				\tmyone = \Qy + 0.35 + 2.85 - \offsety;
				\tmxtwo = \Qx + 4.2 + \offsetx + 0.15 + \offsetextra;
				\tmytwo = \wy + 0.55 - \offsety - \offsetextra;
				\pvxone = \px + 0.75 +\offsetx;
				\pvyone = \py + 0.25-\offsety;
				\pvxtwo = \Qx + 4.65+\offsetx + +\offsetextra;
				\pvytwo = \wy+0.55-\offsety-\offsetextra;
				\outxone= \Qx+4.35+\offsetx+0.15+\offsetextra;
				\outyone = \wy + 0.4-\offsety-\offsetextra;
				\outxtwo = \ox+0.21 + \offsetx;
				\outytwo = \oy+0.1+\offsety;
			}

			\draw [draw=lightgray, fill=lightgray] (\oponex-\opwidth,\oponey-\opwidth) rectangle (\oponex+\opwidth,\oponey+\opwidth) node[pos=0.5] {\tiny{$\bullet$}};

			\draw[{Square[width=0.6mm,length=0.6mm,fill=red]}-,line width=0.15mm] (\tmxone-0.5 + 2* \offsetx, \tmyone) -- (\tmxone, \tmyone)  -- (\tmxone, \tmytwo) -- (\tmxtwo, \tmytwo);

			\draw[{Square[width=0.6mm,length=0.6mm,fill=red]}-,line width=0.15mm] (\pvxone, \pvyone) --  (\pvxone,\pvytwo) -- (\pvxtwo, \pvytwo);

			\draw[-{Square[width=0.6mm,length=0.6mm,fill=red]},line width=0.15mm] (\outxone, \outyone) -- (\outxone, \outyone-0.4) -- (\outxtwo, \outyone-0.4) -- (\outxtwo, \outytwo);

		}

		\draw[{Square[width=0.6mm,length=0.6mm,fill=red]}-,line width=0.15mm] (\ox+9.06, \oy+0.21) -- (\ox+9.3, \oy+0.21);
		\draw[{Square[width=0.6mm,length=0.6mm,fill=red]}-,line width=0.15mm] (\ox+8.96, \oy+0.08) -- (\ox+9.3, \oy+0.08);
		\draw [draw=lightgray] (\ox+9.3,\oy) rectangle (\ox+9.6,\oy+0.3) node[pos=0.5] {\tiny{$\circ$}};
		\draw[->,line width=0.15mm] (\ox+9.6, \oy+0.14) -- (\ox+9.8, \oy+0.14);

		\draw [draw=none] (\ox+9.8,\oy) rectangle (\ox+10.1,\oy+0.3) node[pos=0.5] {
			\scriptsize{\, \, $\preorderPartial_{\currnode \rate \column}$}
		};

		\draw [draw=lightgray, fill=lightgray] (\Qx-1.1, \Qy-2.75) rectangle (\Qx-0.8, \Qy-3.05) node[pos=0.5] {\tiny{$\bullet$}};
		\node [right] at (\Qx-0.9, \Qy-2.95) {\scriptsize{Inner product}};

		\draw [draw=lightgray] (\Qx+1.5, \Qy-2.75) rectangle (\Qx+1.8, \Qy-3.05) node[pos=0.5] {\tiny{$\circ$}};
		\node [right] at (\Qx + 1.7, \Qy-2.95) {\scriptsize{Element-wise multiplication}};

		\draw [fill=SeaGreen] (\Qx+6, \Qy-2.75) rectangle (\Qx+6.3, \Qy-3.05);
		\node [right] at (\Qx + 6.2, \Qy-2.95) {\scriptsize{ShM}};

		\draw [fill=Apricot] (\Qx+7.5, \Qy-2.75) rectangle (\Qx+7.8, \Qy-3.05);
		\node [right] at (\Qx + 7.7, \Qy-2.95) {\scriptsize{Single-thread operations}};

	\end{tikzpicture}
	\caption{Parallel thread-block design to compute post-order partial likelihood vectors $\postorderPartial_{\currnode \rate \column}$.  One block evaluates column block size $(\text{CBS}) \times \states$ entries in parallel and prefetches pruning block size $(\text{PBS}) \times \states$ transition probability entries at time within an inner serial loop.}
	\label{fig:postOrderKernel}
\end{figure}

\begin{figure}[H]
	\center
	\includegraphics[width=\textwidth]{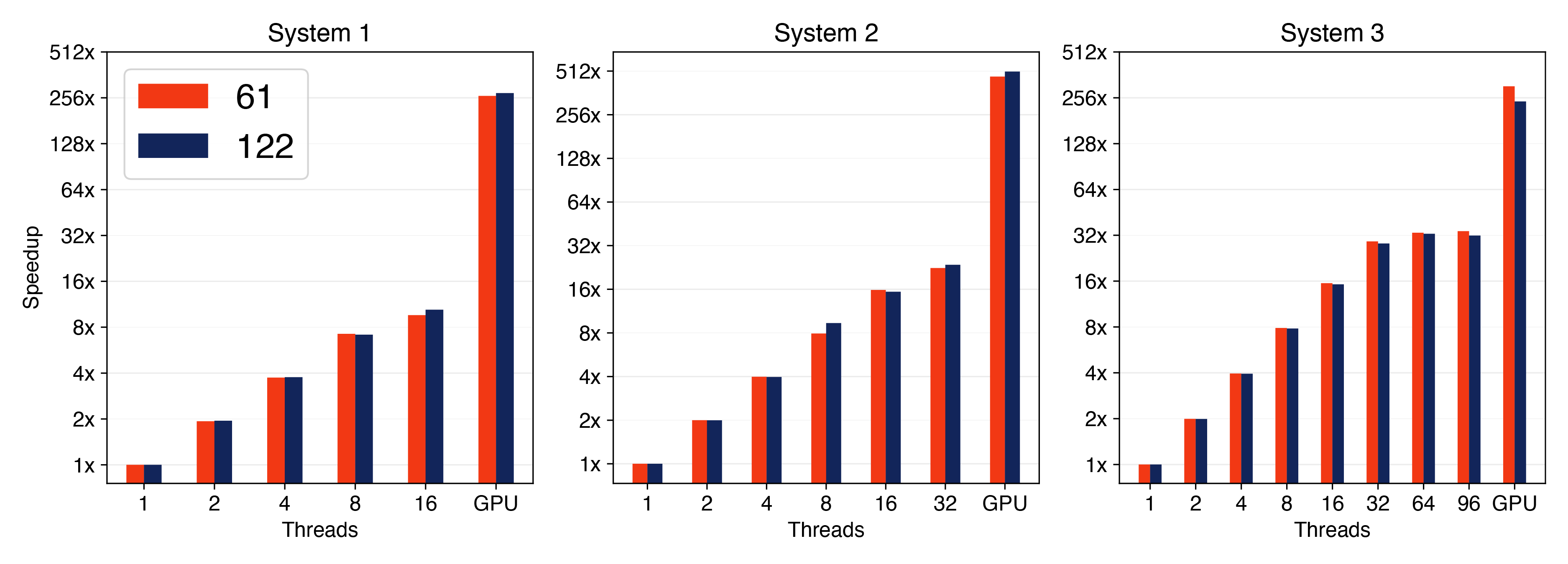}
	\caption{Speedup of GPU and multi-core CPU instances over a single CPU thread for five MCMC iterations to infer branch-specific evolutionary rates using the yeast dataset for a Yang codon model with a state-space size of 61 and a MMM model with a state-space size of 122. Speedup factors are reported on a log-scale.}
	\label{figure:statespace61vs122}
\end{figure}

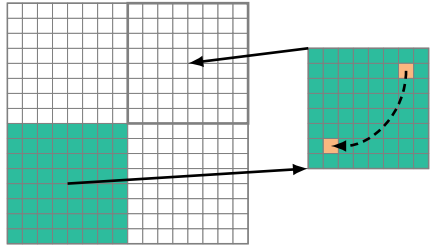
\begin{figure}[H]
  \centering
  \newcommand\ox{0}
  \newcommand\oy{0}
  \newcommand\cellwidth{0.2}
  \begin{tikzpicture}

	\grid{\ox}{\oy}{16}{16}{\cellwidth}{-1}{-1}{8}{9}
	 \draw[draw=gray,fill=none,line width=1pt](\ox+1.8,\oy+1.8) rectangle (\ox+3.4, \oy+3.4);

    \grid{\ox+4}{\oy+1}{8}{8}{\cellwidth}{-1}{-1}{8}{9}
    \draw[draw=gray,line width=0.1mm, fill=Apricot] (\ox+4+0.4, \oy+1+0.4) rectangle (\ox+4+0.6, \oy+1+0.6);
    \draw[draw=gray,line width=0.1mm, fill=Apricot] (\ox+4+1.4, \oy+1+1.4) rectangle (\ox+4+1.6, \oy+1+1.6);

    \draw [-latex,line width=1pt] (\ox+4.2,\oy+2.8) -- (\ox + 2.6, \oy + 2.6);
    \draw [-latex,line width=1pt] (\ox + 1, \oy + 1) -- (\ox+4.2,\oy+1.2);

    \draw [-latex,black,dash pattern=on 3pt off 2pt, line width=1pt] (\ox+4+1.5, \oy+1+1.5) to [out=-90,in=0] (\ox+4+0.5, \oy+1+0.5);

  \end{tikzpicture}
  \caption{Parallel thread-block design to calculate the transpose of the finite-time transition probability matrix. One block calculates the transpose of MULTIPLY\_BLOCK\_SIZE $\times$ MULTIPLY\_BLOCK\_SIZE tile of the matrix.}
  \label{fig:matrixtranspose}
\end{figure}

\renewcommand{\thealgorithm}{S1}

  \begin{algorithm}
    \caption{GPU-based parallel computation of the finite-time transition probability matrix transpose.}
    \label{alg:matrixtranspose}
    \begin{algorithmic}[1]
      \State \textbf{define} MULTIPLY\_BLOCK\_SIZE (MBS) $= $ number of rows and columns processed in parallel per thread-block.
      \For {\textbf{all} thread-blocks ($\text{tile-row} = 1, \ldots , \lceil \states/\text{MBS} \rceil$ and $\text{tile-col}=1, \ldots , \lceil \states/\text{MBS} \rceil$) \textbf{in parallel}}
      	\For {\textbf{all} threads in block ($\state = 1, \ldots, \text{MBS}  \text{ and } \estate = 1, \ldots, \text{MBS}$) \textbf{in parallel}}
      		\State A[\state][\estate] $\gets \probabilityEntry{r}{\branchLength_{\currnode}}{\state}{\estate}$
      		\State return A[\estate][\state]
      	\EndFor
      \EndFor
    \end{algorithmic}
  \end{algorithm}

\end{document}